\documentclass[amsmath,amssymb,prd,aps,psfig,floats]{revtex4}
\usepackage{bm}
\usepackage{graphicx}
\usepackage{times}
\newcommand{\EQ}{\begin{equation}}
\newcommand{\EN}{\end{equation}}
\newcommand{\EQA}{\begin{eqnarray}}
\newcommand{\ENA}{\end{eqnarray}}
\newcommand{\eq}[1]{(\ref{#1})}

\newcommand{\Eq}[1]{Eq.~(\ref{#1})}

\newcommand{\Sec}[1]{Section~\ref{#1}}

\newcommand{\App}[1]{Appendix~\ref{#1}}

\newcommand{\bra}[1]{\langle #1\rangle}

\newcommand{\kkk}{\hat{\bm{k}}}

\newcommand{\xx}{\bm{x}}

\newcommand{\vv}{\bm{v}}

\newcommand{\mm}{\bm{m}}

\newcommand{\VV}{\bm{V}}

\newcommand{\nn}{\bm{n}}

\newcommand{\ff}{\bm{f}}

\newcommand{\FF}{\bm{F}}

\newcommand{\kk}{\bm{k}}

\newcommand{\nab}{\bm{\nabla}}

 {}

\newcommand{\ii}{\mathrm{i}}

\newcommand{\T}{\,{\rm T}}

\newcommand{\K}{\,{\rm K}}

\newcommand{\Mpc}{\,{\rm Mpc}}

\begin{document}
\draft
\preprint{} 
\title{The Physics of CMBR Anisotropies}
\author{Kandaswamy Subramanian}
\affiliation{Inter University Centre for Astronomy and Astrophysics, Post Bag 4,
Ganeshkhind, Pune 411 007, India. }
\begin{abstract}
The observed structures in the universe are thought to have arisen
from gravitational instability acting on small fluctuations
generated in the early universe. These spatial fluctuations are imprinted
on the CMBR as angular anisotropies. 
The physics which connects initial fluctuations in the early 
universe to the observed anisotropies is fairly well understood,
since for most part it involves linear perturbation theory.
This makes CMBR anisotropies one of the cleanest probes of the 
initial fluctuations, various cosmological parameters
governing their evolution and also the geometry of the
universe. We review here in a fairly pedagogical manner 
the physics of the CMBR anisotropies and explain
the role they play in probing cosmological parameters,
especially in the light of the latest observations from the WMAP 
satellite.

\end{abstract}

\date{\today}
\maketitle
\renewcommand{\thefootnote}{\arabic{footnote}} \setcounter{footnote}{0}

\section{ Introduction}

The cosmic microwave background radiation (CMBR) 
is of fundamental importance in
cosmology. Its serendipitous discovery by Penzias and Wilson \cite{PW65},
gave the first clear indication of an early hot 'Big bang" stage on 
the evolution of the universe. The subsequent verification by host
of experiments, culminating in the results of the COBE 
satellite confirmed that its spectrum 
is very accurately Planckian \cite{mather_etal},
with a temperature $T= 2.725$. This is the firmest evidence
that the universe was in thermal equilibrium at some early stage. 
Indeed the observed limits on the spectral distortions severely
constrain any significant energy input into the CMBR 
below $z < 10^7$ or so \cite{ydistortion}. 

Shortly after its discovery, it was also predicted
that the CMBR should show angular variations in its temperature,
due to photons propagating in an inhomogeneous universe \cite{SW67}.
In the standard picture, the baryonic matter in the
early universe was in a highly ionized form with radiation
strongly coupled to the baryons. As the universe expanded, the matter cooled
and atoms formed below about $3000^o$ K. After this epoch the photon
mean free path increased to greater than the present Hubble radius, and
they could free stream to us. These are the photons that we detect
in the CMB. They carry information both about the conditions at the
epoch of their last scattering, as well as processes which affect their
propagation subsequently. Fluctuations in the early universe
result in inhomogeneities on the 'last scattering surface' (henceforth LSS).
These inhomogeneities should be seen today as angular anisotropies in the
temperature of the CMB. Further, the CMB photons are
influenced by a number of gravitational and scattering effects
during their passage from the LSS to the observer. These are also
expected to generate additional CMBR anisotropies.

These CMBR anisotropies took a long time to be discovered and its
absence in the early observations were beginning to prove
embarrassing, for theories of structure formation. 
It was not until 1992 that the temperature anisotropies
in the CMBR were detected, on large angular scales, by the
Differential Microwave Radiometers (DMR) experiment on
the COBE satellite \cite{smoot92}. 
The fractional temperature anisotropies are 
at the level of $10^{-5}$ and ruled out some of the earlier baryon
dominated models, and hot dark matter dominated models, 
but were quite consistent with expectations from
latter Cold Dark matter models of structure formation \cite{EBW92,PN92}.

Since the COBE discovery a large number of expreiments have 
subsequently probed the CMBR angular anisotropies over a variety of
angular scales, from degrees to arc minutes (cf. \cite{bcp03} for
a recent review). This has culminated in the release of the first year
all-sky data from the Wilkinson Microwave Anisotropy Probe (WMAP)
satellite \cite{bennett03}. These observations, especially the
'acoustic oscillations' which are inferred from the
anisotropy power spectrum, have led to the confirmation of a popular 
'standard' picture for structure formation; one where an early
epoch of inflation generated adiabatic perturbations
in a spatially flat universe. The observed anisotropy patterns
also allow cosmological parameters to be probed with 
considerable precision, especially when combined with other data sets
related to the observed inhomogeneous universe \cite{spergel03,tegmark04}. 
It has therefore become imperative for the modern cosmologist to understand 
the physics behind CMBR anisotropies.  We review here  
in a pedagogical fashion, the relevant physics of the temperature 
anisotropies and also briefly mention the polarization of the CMBR. 
There are a large number of reviews \cite{cmbrev,hudodr,hurev,zal03}, and
text books \cite{paddycos,dodelson} on this subject. The authors aim is to
present some of these ideas in a manner in which he, as a non expert, 
understood the subject, which may be of use to some!

\section{ The CMB observables}

The CMB is described by its brightness (or intensity)
distribution. Since the spectrum of the CMB brightness, seen 
along any direction on the sky ${\bf n}$, is very close to 
thermal, it suffices in most cases to give the temperature $T({\bf n})$.
The temperature is very nearly uniform
with fluctuations $\Delta T(\nn)$ at the level of
$10^{-5} T$, after removing a dipole contribution.
It is convenient to expand the temperature anisotropies
$\Delta T(\nn)/T = \Theta(\nn)$ at the observer in 
spherical harmonics
\EQ
\Theta({\bf n}) \equiv {\Delta T\over T}(\theta,\phi)
= \sum_{lm} a_{lm} Y_{lm}(\theta,\phi) 
\label{alm}
\EN
with $a_{lm}^* = (-1)^m a_{l-m}$, since the temperature is a real
quantity.

In the standard picture, the universe is assumed to have
evolved from density fluctuations initially described by a
Gaussian random field, and one can then 
take $\Theta$ to be a Gaussian random field.
In this case $a_{lm}$'s are also Gaussian random variables
with zero mean and a variance completely described by their power spectrum,
\EQ
\bra{a_{lm}a_{l'm'}} = C_l\delta_{ll'}\delta_{mm'}.
\label{cldef}
\EN
Here we have assumed also the statistical isotropy of $\Theta(\nn)$ field
because of which the power spectrum is independent of $m$.
Theoretical predictions of CMBR anisotropy are then compared
with observations by computing the $C_l$'s or the
correlation function $C(\alpha) = \bra{\Theta(\nn)\Theta(\mm)}$, 
where if we have statistical isotropy, $C$ depends only on
$\cos\alpha = \nn\cdot\mm$.
From \Eq{cldef} and the addition theorem for the spherical
harmonics, we have
\EQ
C(\alpha)=  \sum_{lm} \sum_{{l^\prime} {m^\prime}} \bra{a_{lm}
a^*_{{l^\prime} {m^\prime}}}Y_{lm} Y^*_{{l^\prime } {m^\prime }}
= \sum_l C_l {2l +1\over 4\pi} P_l(\cos\alpha).
\label{cobs}
\EN
The mean-square temperature anisotropy, 
$\bra{(\Delta T)^2} = T^2 C(0)$ is  
\EQ
\bra{(\Delta T)^2} = T^2\sum_l C_l {2l +1\over 4\pi} \approx T^2
\int \frac{l(l+1)C_l}{2\pi} \ d \ln l
\EN
with the last approximate equality valid for large $l$, 
and so $l(l+1)C_l/ 2\pi$ is a measure of the power
in the temperature anisotropies, per logarithmic
interval in $l$ space. (We will see below that
scale invariant potential perturbations
generate anisotropies, due to the Sachs-Wolfe effect \cite{SW67}, 
with a constant $l(l+1)C_l$,
which provides one motivation for this particular combination).

Note that the CMB brightness and hence $\Theta$ is also a function
of the space-time location $({\bf x} ,\eta)$ of the observer.
Here $\xx$ is the conformal spatial and $\eta$ the conformal
time co-ordinates respectively (see below).
One computes the correlation function $C(\alpha)$
predicted by a given theory by taking the ensemble average of 
$\bra{\Theta({\bf x_0},\eta_0,\nn)
\Theta({\bf x_0},\eta_0,\mm)}$. For the statistically isotropic case
this again only depends on $\cos\alpha = {\bf n}\cdot{\bf m}$.
Further the Fourier component
of $\Theta$, for every $\kk$ mode, 
often depends on $\nn$ only through $\kkk\cdot\nn = \mu$, where
$\kkk = \kk/\vert\kk\vert$. One can then conveniently expand
$\Theta$ in a Fourier, Legendre series as
\EQ
\Theta({\bf x_0},\eta_0,{\bf n}) =  
\int {d^3{\bf k}\over (2\pi)^3 } e^{i{\bf k}.{\bf x_0}} 
\sum_l (-i)^l (2l +1) a_l(\kk,\eta_0) P_l(\kkk\cdot\nn)
\label{aldef}
\EN
For a homogeneous, isotropic, Gaussian random $\Theta$ field,
$\bra{a_l({\bf k},\eta_0)a^*_{l^\prime}
({\bf p},\eta_0)}
= \bra{\vert a_l(k,\eta_0)\vert^2} \delta_{l,l^{\prime}}
(2\pi)^3 \delta^3({\bf k} -{\bf p})
$, where the power spectrum $\bra{\vert a_l(k,\eta_0)\vert^2}$ 
depends only on $k=\vert\kk\vert$.
One then gets
\EQ
C(\alpha) 
= \sum_l \frac{2}{\pi} \int {dk \over k} k^3 \bra{\vert a_l(k,\eta_0)\vert^2}
\frac{2l+1}{4\pi} P_l(\cos\alpha),
\label{cofun}
\EN
where we have used the addition theorem 
\EQ
P_l(\kkk\cdot\nn) = \frac{4\pi}{2l+1} \sum_m Y_{lm}(\nn)Y^*_{lm}(\kkk)
\EN
and carried out the angular
part of the integral over $d^3{\bf k}$ using the orthogonality
of the Spherical harmonics. Comparing \Eq{cofun} with \Eq{cobs}
we see that
\EQ
C_l = \frac{2}{\pi} \int {dk \over k} k^3 \bra{\vert a_l(k,\eta_0)\vert^2}
\label{alcl}
\EN
We will use this equation below to calculate $C_l$'s 
for various cases. One can roughly set up a correspondence between 
angular scale at the observer $\alpha$, the corresponding  $l$ value
it refers to in the multipole expansion of $\bra{\Theta^2}$ and also 
the corresponding co-moving wavenumber $k$. One has 
$(\alpha/1^{o}) \approx (100/l)$ and $l \approx kR^*$ where
$R^*$ is the comoving angular diameter distance to
the LSS and is $\sim 10h^{-1}$ Gpc,
for a standard $\Lambda$CDM cosmology (see below).

We show in Figure 1 a plot of the temperature anisotropy
$\Delta T = T \sqrt{l(l+1)C_l/2\pi}$ and polarization anisotropy 
versus $l$, for a standard
$\Lambda$CDM cosmology, got by running the publicly available code 
CMBFAST \cite{cmbfast}. One sees a number of features in such a plot, 
a flat plateau at low $l$ rising to several peaks and dips, 
as well as a cut-off at high $l$. Our aim will be
to develop a physical understanding of the various features that
the figure displays. We now turn to the formalism for computing 
the $C_l$`s for any theory.

\section{ The Boltzmann equation}

The distribution function for photons, $f(x^{i},p_{\beta})\equiv 
f({\bf x},{\bf p},\eta)$,
is defined by giving their number in an infinitesimal phase space volume,
$dN = f(x^{\alpha},p_{\beta},\eta)d^3x^{\alpha} d^3p_{\beta}$.
Note that we will use Greek letters for purely spatial 
co-ordinates and Latin ones for space-time co-ordinates. 
We can write $dN$ in a co-ordinate independent way,
\EQ
dN = \int_{p_0} \frac{d^4p_k}{\sqrt{-g}} \ f(x^i,p_j) \  
p^l d\Sigma_l  \ 2\delta[p_mp^m]
\label{dn}
\EN
where $g$ is the determinant of the metric, 
$d\Sigma_i= \sqrt{-g} \ \epsilon_{ijkl} \ [dx^i\wedge dx^j\wedge dx^k]/3!$ 
is an infinitesimal spacelike hypersurface, $p_i$ the photon 4-momentum and the
delta function ensures that $p_i$ is a null vector. 
To get the simple expression for $dN$, one chooses a time slicing with
$d\Sigma_i \equiv (\sqrt{-g}dx^1dx^2dx^3,0,0,0)$ and carries out
the integral over $p_0$ retaining
only the positive energy part of the delta function. 
This shows explicitly that
$f$ is a co-ordinate invariant scalar field.
Further, in the absence of collisions,  
both $dN$ and the phase-space volume $d^3x^{\alpha} d^3p_{\beta}$ would
be conserved along the photon trajectory and hence also the phase space
density $f$. If $\lambda$ is an affine parameter along the 
null geodesic, we will then have $df/d\lambda =0$. On the other hand when 
collisions are present the distribution function will change. This situation 
is generally handled by introducing a collision term on the RHS of the CBE, 
that is writing $df/d\lambda = \tilde c(f)$. Further it is generally 
convenient to use the time co-ordinate itself, say $\eta$, 
as the independent parameter along the photon trajectory and write 
$df/d\lambda = (d\eta/d\lambda) (df/d\eta) 
=\tilde c(f) = (d\eta/d\lambda) c(f)$. One then has
\EQ
{df\over d\eta} = {\partial f\over \partial \eta} + {dx^{\alpha}\over d\eta}
{\partial f\over \partial x^{\alpha}} + {dp_{\alpha}\over d\eta}
{\partial f\over \partial p_{\alpha}} = c(f)
\label{bolti}
\EN
We look at this equation in the spatially flat, perturbed FRW universe. 
Its metric in the conformal-Newton gauge is
\EQ
ds^2 = a^2(\eta)\left[(1 + 2\phi)d\eta^2 - (1-2\phi)(dx^2 + dy^2 + dz^2)\right]
\label{metric}
\EN
Here $a(\eta)$ is the expansion factor and $\eta$ is the conformal
time, related to the proper time by $a(\eta) d\eta = dt$. 
(We adopt $c=1$ units). We have assumed that
a single potential $\phi$ describes the scalar perturbation,
which holds if the source of the perturbations is a perfect fluid 
with no off-diagonal components to the energy-momentum tensor.

Since the photon 4-momentum $p^i$ is a null vector
we have $ g^{ik} p_ip_k =0$. We choose the
photon 4-momentum to have components
$p_i \equiv (\sqrt{g^{\alpha\alpha}/g^{00}} \ p, -pn^\alpha)$
where, we have defined 
the magnitude of the
spatial component of (co-variant) momentum,
$p = \Sigma_{\alpha} p_{\alpha} p_{\alpha}$. Also
$\nn$ is a unit vector in the direction of the photon 
3-momentum $p^{\alpha}$.

Then to linear order in the perturbed potential, 
\EQ
\frac{dx^{\alpha}}{d\eta} 
= {dx^{\alpha}/d\lambda\over d\eta/d\lambda} 
= {p^{\alpha}\over p^0} = n^{\alpha}(1 + 2\phi). 
\label{vel}
\EN
The geodesic equation for the photons to the linear 
order in the perturbations is 
\EQ
{dp_i\over d\eta} = 2 p{\partial \phi \over \partial x^i}.
\label{gmom}
\EN
The Boltzmann equation then becomes
\EQ
{df\over d\eta} = {\partial f\over \partial \eta} +(1+ 2\phi) n^{\alpha}
{\partial f\over \partial x^{\alpha}} + 
2p{\partial \phi\over \partial x^{\alpha}} 
{\partial f\over \partial p_{\alpha}} = c(f)
\label{bolt}
\EN

An observer at rest in the perturbed FRW
universe, has a 4-velocity $u^i \equiv (1/\sqrt{g_{00}}, 0)$.
So the energy of the photon measured by such an observer
is $E = p_iu^i = p(1+\phi)/a$. In the unperturbed FRW
universe, the energy simply redshifts with expansion with
$E = p/a$. The distribution function for the photons, 
in the absence of perturbations
is then described by the Planck law, 
\EQ
f_b\left(\frac{p}{T}\right) = \frac{A}{exp(p/T) - 1} .
\label{pla}
\EN
Defining the perturbed phase space density $f_1(\xx, p, \nn, \eta) =
f(\xx, p, \nn, \eta) - f_b(p)$, to
linear order the Boltzmann equation becomes
\EQ
{\partial f_1\over \partial \eta} +\nn\cdot\nab f_1
-2p \ \nn\cdot\nab\phi
{\partial f_b\over \partial p} = c(f)
\label{boltl}
\EN
where we have replaced $f$ by $f_b$ in the term last on the LHS of
\Eq{bolt} and used $(\partial f_b/\partial p_{\alpha}) = (p_{\alpha}/p)
(\partial f_b/\partial p) =-n^{\alpha} (\partial f_b/\partial p)$.

In the perturbed FRW universe we note that both the perturbed
trajectory and the effect of collisions (under the Thomson scattering
approximation) do not depend on the photon energy. This
motivates us to define the perturbed phase space density
in terms of a purely temperature perturbation $\delta T(\xx,\eta,\nn)$ 
in $f_b$. We take
\EQ
f({\bf x},p,{\bf n},\eta) = 
f_b\left[ \frac{p}{T + \delta T(\xx,\eta,\nn)}\right].
\label{tdef}
\EN
(Note that in such an approximation, we are also neglecting
the effects of any spectral distortion).
To the first order in $\delta T/T$ one can expand $f$ in \Eq{tdef} to get
$f_1 = -p (\partial f_b/\partial p) (\delta T/T)$.
Again because both the perturbed
trajectory and the effect of collisions 
do not depend on $p$, we will usually
integrate over $p$'s. It is then useful to deal
with not the full phase space density, but just its
associated fractional brightness perturbation
defined by
\EQ
i = \frac{\int f_1p^3dp}{ \int f_bp^3dp} 
= 4 \frac{\delta T}{T}(\xx,\eta,\nn).
\label{brit}
\EN
To appreciate better the meaning of $(\delta T/T) = i/4$ let
us look at the energy momentum tensor of the photons.
This is 
\EQ
T^i_j = \int {d^4p_k \over \sqrt{-g} } \  p^i p_j 2 \ \delta(p^mp_m) f 
= \int {d^3p_{\alpha} \over \sqrt{-g} } {p^i p_j \over p^0} f
\label{enmom}
\EN
The energy density in the perturbed FRW universe, with metric given by 
\Eq{metric} is
\EQ
\rho=T^0_0 = {(1 + 4\phi)\over a^4} \int p^3 dp \  d\Omega \ f
= \rho_R(1 + 4\phi) 
\left[1 + \int i {d\Omega \over 4\pi} \right]
\label{rh}
\EN
where $\rho_R = (4\pi/a^4) \int p^3dp f_b(p)$ 
is the radiation energy density in the absence of perturbations.
Let us define $i_0 = \int i (d\Omega/ 4 \pi)$ and $(\delta T/T)_0 = 
\int (\delta T/T) (d\Omega/ 4 \pi)$, the zeroth moments of the 
perturbed brightness $i$ and temperature $(\delta T/T)$ respectively,
over the directions of the photon momenta.
The fractional perturbation to the radiation energy density is given by
\EQ
\delta_R = {\rho -\rho_R \over \rho_R} = 4\phi + i_0
= 4\left[ \left( \frac{\delta T}{T}\right)_0 + \phi \right]
\label{pert}
\EN
Note that in the conformal Newton gauge the radiation density 
perturbation has an additional contribution from the perturbed potential itself
(over and above the contribution from the perturbed distribution function).
One may feel that this differs from the naive expectation that
the radiation energy
density to vary as $\rho \propto T^4$ and hence the 'physical'
temperature perturbation go as just $(1/4) \delta \rho/\rho$.
Since the energy of a photon seen by an observer at rest in the
perturbed FRW universe is $E = p(1 +\phi)/a$ (see above),
one can write the phase space density in the perturbed FRW universe,
to linear order as
\EQ
f = f_b\left(\frac{p}{\T + \delta T}\right)
= f_b\left(\frac{Ea}{(\T + \delta T)(1 +\phi)}\right)
= f_b\left(\frac{Ea}{\T + (\delta T +\phi)}\right).
\label{defthet}
\EN
This shows that $\Theta = \delta T +\phi$ is indeed the 
'physical' temperature perturbation measured
by an observer at rest in the perturbed FRW universe and
that $\Theta_0 = \int \Theta (d\Omega/4\pi) = \delta_R/4$,
as expected.

\section{ The collision term and the equation for the perturbed brightness}
\label{boltz}

We now consider the effect of collisions.
The process that we wish to take into account is the scattering 
between photons and electrons.
In fact to linear order it is sufficient to consider the 
Thomson scattering limit of negligible energy transfer in the 
electron rest frame. Since the distribution function is a scalar,  
the effects of collision are most simply calculated by going to the 
electron (or fluid) rest frame and transferring the results to any 
other co-ordinate frame. Suppose the distribution function changes 
by $ df = d\bar f= d\tau \bar c$ in the fluid rest frame.
Henceforth quantities with an 'overbar' will represent variables in the 
fluid rest frame. Then one can write 
$df/d\eta = (d\tau/d\eta) \bar c$. 
The differential cross section for Thomson scattering of unpolarized radiation
is given by
\EQ
\frac{d\sigma}{d\Omega'} = 
\frac{\sigma_T}{4\pi} \left( 1 + \frac{P_2(\bar\nn\cdot\bar\nn')}{2} \right)
\label{thom}
\EN
where $\sigma_T$ is the Thomson cross section
and $\bar{\bf n}$ and $\bar{\bf n}^{\prime}$ are the unit vectors specifying
the direction of the initial and the scattered photon momenta in the 
fluid rest frame.
The collision term $df/d\tau = d\bar f/d\tau = \bar c$ will have a source due
to the photons scattered into the beam from
a direction $\bar\nn'$ and a sink due to scattering out of the beam. 
So we have
\EQ
\bar c(\bar f) 
= n_e \sigma_T \int \frac{d\Omega^{\prime}}{4\pi}
\left[1 +\frac{P_2(\bar\nn\cdot\bar\nn')}{2}\right]
\left[\bar f(\bar p,\bar\nn')-\bar f(\bar p,\bar{\bf n})\right]
\label{dffir}
\EN
where the integration over $d\Omega^{\prime}$ is 
over the directions of $\bar{\bf n}^{\prime}$.

In order to derive the equation satisfied by the brightness perturbation we 
multiply the linearized Boltzmann equation \eq{boltl}  
by $p^3$ and integrate over $p$ to get 
\EQ
{\rho_R a^4\over 4\pi}\left[ {\partial i\over \partial \eta} +{\bf n}.
\nabla i  + 8{\bf n}.\nabla \phi\right] = {d\tau \over d\eta} 
\int p^3 dp \  \bar c(\bar f) \label{btli}
\EN
We simplify the collision term on the RHS of
\Eq{btli}, in \App{collision}.
From \Eq{btli}, \Eq{ifrt} and \Eq{atwo} the equation 
satisfied by $i$, to the leading order of the perturbations, 
is given by
\EQ
{\partial i\over \partial \eta} +{\bf n}.
\nabla i  + 8{\bf n}.\nabla \phi = 
n_e \sigma_T a \left[i_0 +
4{\bf n}.{\bf v} + \frac{1}{10}\sum_m Y_{2m}(\nn)i_{2m} - i \right].
\label{final}
\EN
The effects of Thomson scattering is to drive the photon
distribution such that the RHS of \Eq{final} would vanish.
If the scattering cross section had been isotropic, then
$i$ would have been driven to $i_0$ in the fluid rest frame;
but in the frame where the fluid moves there is a doppler shift.
In addition the anisotropy of Thomson scattering introduces
the dependence on the quadrupole moment of the brightness.
The perturbed brightness equation will be used to derive the
equation for the CMBR anisotropies, and also the dynamics
for the baryon photon fluid. 

\section{ Integral solution for CMBR anisotropies}

Consider the perturbed brightness equation \Eq{final} in
Fourier space, in terms of the Fourier coefficients
$\hat\Theta(\kk,\eta,\nn)$ of the 'physical'
temperature perturbation $\Theta = \delta T/T + \phi = i/4+\phi$, that is
\EQ
\Theta(\xx,\eta,\nn) = \int \frac{d^3\kk}{(2\pi)^3} e^{\ii \kk\cdot\xx}
\hat\Theta(\kk,\eta,\nn).
\label{thetahat}
\EN
Henceforth we shall denote the Fourier transform of 
any quantity $A$ by $\hat A$, except for
the velocity ($\vv$) and potential ($\phi$) whose Fourier transforms
are denoted by $\VV$ and $\Phi$ respectively.
We also assume that these Fourier co-efficients
depend on $\nn$ only through $\mu = \kkk\cdot\nn$, as will
obtain for example for scalar perturbations. In this case
one can choose an axis for each $\kk$ mode
such that $i_{2m} = 0$ for $m \ne 0$ and 
$\sum_m Y_{2m}(\nn)\hat i_{2m}/10 = -\hat \Theta_2 P_2(\mu)/2$,
where 
\EQ
\Theta(\kk,\eta,\mu) = \sum_l (-\ii)^l (2l+1)\hat 
\Theta_l(\kk,\eta) P_l(\mu)
\label{thetl}
\EN
is the expansion of $\hat\Theta$ in a Legendre series.
For scalar perturbations, $\VV$ is also parallel to
the $\kk$ vector, and so $\nn\cdot\VV = (\nn\cdot\kkk) V
= V\mu$. Further, it is much more convenient to work with
the equation for the combination $\hat\Theta + \Phi$.
From the Fourier transform of \Eq{final} we then have
\EQ
(\dot{\hat\Theta} +\dot\Phi) 
+(\ii k\mu +n_e\sigma_Ta)[\hat\Theta + \Phi] = 
n_e\sigma_Ta \ S(\kk,\eta, \mu)
+ 2\dot\Phi,
\label{forfin}
\EN
where, henceforth an over dot will denote a partial derivative, i.e.
$\dot f = \partial f/\partial\eta$, for any $f$. 
We have also defined the source function
\EQ
S(\kk,\eta, \mu) = 
\left[\hat\Theta_0 + \Phi +V \mu  
-\frac{P_2(\mu)\hat\Theta_2}{2}\right]
\label{source}
\EN
Suppose we define the differential optical 
depth to electron scattering $d\tau = n_e \sigma_T a \ d\eta =
n_e\sigma_T  dt$. Then one can solve \Eq{forfin} formally as
\EQA
\hat\Theta(\kk,\eta_0,\mu) + \Phi(\kk,\eta_0) &=& 
[\hat\Theta(\eta_i) + \Phi(\eta_i)]
e^{-ik\mu(\eta_0 -\eta_i)} e^{-\tau(\eta_0,\eta_i)} 
\nonumber \\
&+&
\int_{\eta_i}^{\eta_0} d\eta \ e^{- \tau(\eta_0,\eta)} 
\left[\dot\tau
S({\bf k}, \eta^{\prime}, {\bf n}) + 2\dot\Phi \right] 
e^{-ik\mu(\eta_0-\eta)}
\label{Isol}
\ENA
where we have defined the optical depth to electron
scattering between epochs $\eta$ to the present $\eta_0$
\EQ
\tau(\eta_0,\eta) = \int_{\eta}^{\eta_0} d\eta' n_e(\eta') \sigma_T a(\eta'),
\label{optical}
\EN
and $\dot\tau(\eta) =d\tau(\eta,\eta')/d\eta = n_e(\eta)\sigma_T a(\eta)$.
The first term on the RHS of \Eq{Isol} can be neglected by taking a small 
enough initial time $\eta_i$, because the exponential damping for large 
optical depths. One can then simplify \Eq{Isol} to get at 
the present epoch $\eta_0$
\EQ
\hat\Theta(\kk,\eta_0,\mu) + \Phi(\kk,\eta_0)
= \int_{\eta_i}^{\eta_0} d\eta \ 
S(\kk,\eta,\mu ) g(\eta_0,\eta) e^{ -ik\mu(\eta_0-\eta)}
+2 \int_{\eta_i}^{\eta_0} d\eta
\ \dot\Phi \
e^{ -ik\mu(\eta_0-\eta)}
e^{-\tau(\eta_0,\eta)}.
\label{ISOL}
\EN
We have defined above the visibility function
\EQ
g(\eta_0,\eta) = \dot\tau (\eta) e^{-\tau(\eta_0,\eta)} =   
n_e(\eta) \sigma_T a(\eta) \ \exp\left[-\int_{\eta}^{\eta_0}
d\eta' n_e(\eta') \sigma_T a(\eta')\right],
\label{visfn}
\EN
such that $g(\eta_0,\eta)d\eta$  
gives for every $\eta_0$ the probability that the last scattering of 
a photon occurred  in the interval  $(\eta, \eta + d\eta)$. 
Suppose $\eta_0$ is the conformal time at the present epoch. Then as 
$\eta$ decreases from $\eta_0$, the optical depth to electron 
scattering will increase and 
so will $g$. However far back into the past when $\tau \gg 1$, $g$ will 
be exponentially damped. So the visibility function generally increases 
as one goes into the past attains a maximum at an 
'epoch of last scattering' and decreases exponentially thereafter. 
Its exact behavior of course depends on the evolution of the 
free electron number density during the recombination epoch and
also on the subsequent ionization history of the universe. 
If the universe went through a standard recombination 
epoch with no significant reionization thereafter, then the 'surface of 
last scattering' is centered at $z \approx 1100$ with a very small half width 
$\Delta z \approx 100$. If on the other hand the universe got 
significantly reionized at high redshifts, as it seems to be indicated
by the WMAP observations, some fraction $\tau_{ri} \sim 0.17$,
of the photons will suffer last scattering surface at later epochs.

We can calculate $a_l(\kk,\eta_0)$ by taking the Legendre
transform of both sides of \Eq{ISOL}. 
Note that the term $\Phi({\bf k},\eta_0)$ on the LHS of \Eq{ISOL},  
does not depend on the photon direction and so does not contribute 
to CMBR anisotropy at all.  
Using the expansion of plane-waves in terms of
spherical waves, 
\EQ
e^{ -ik\mu x} = \sum_l (-\ii)^l (2l+1)j_l(x) P_l(\mu), 
\label{plane}
\EN
and writing
$\mu \exp(-\ii\mu k x) = i d(\exp(-\ii\mu k x))/d(kx)$, we get
\EQA
a_l(\kk,\eta_0) = \int_0^{\eta_0} d\eta && g(\eta_0,\eta)
\left[ (\hat\Theta_0 + \Phi)j_l(k\Delta\eta) +
\ii V j_l'(k\Delta\eta)
+ \frac{\hat\Theta_2}{2} \ \frac{\left(3j_l''(k\Delta\eta) + j_l(k\Delta\eta)
\right)}{2} \right]
\nonumber \\
&&
+ \int_0^{\eta_0} d\eta \ e^{-\tau(\eta_0,\eta)}
2\dot\Phi j_l(k\Delta\eta)
\label{aliso}
\ENA
Here $j_l(kx)$ is the spherical Bessel function, and
$j_l'$ denotes a derivative with respect to the argument,
$\Delta\eta = \eta_0 - \eta$.

Let us interpret the various terms in \Eq{aliso}.  
This equation shows that 
anisotropies in the CMBR result from a combination of
radiation energy density perturbations $\hat\Theta_0$ 
and potential perturbations $\Phi$ (the monopole term),
on the last scattering surface and 
the doppler effect due to the line of sight component
of the baryon velocity $V$ (the dipole term).
The anisotropy of the Thomson scattering cross section also
leads to a dependence on the radiation quadrupole $\hat\Theta_2$.
The spherical Bessel function and its derivatives in front 
of these terms project variations in space, at the conformal
time $\eta$ around last scattering, to the angular (or $l$) 
anisotropies at the present epoch $\eta_0$. (A popular jargon
is to say that the monopole, dipole and quadrupole at last scattering
free stream to produce the higher order multipoles today). 
These spherical Bessel
functions generally peak around $k\Delta\eta \approx l$. 
The multipoles $l$ are then probing
generally spatial scales with wavenumber $k \sim l/\Delta\eta$
at around last scattering. The visibility function weighs
the contribution at any conformal time $\eta$ by the probability
of last scattering from that epoch.
Finally, the last term ($\dot\Phi$ term) shows that any variation
of the potential along the line of sight
will also lead to CMBR anisotropies, and is usually
referred to as the integrated Sachs-Wolfe (ISW) effect.

In the limit of a very narrow LSS at $\eta=\eta^*$,
For angular scales (and $l$'s) such that,
their associated spatial scales at the LSS are much larger than
the LSS thickness, one can take the variation
of the $j_l's$ with $\eta$ to be much slower than that
of the visibility function. In such a narrow LSS
approximation, we get
\EQ
a_l(\kk,\eta_0) = 
(\hat\Theta_0 + \Phi)(\eta^* )j_l(kR^*) +
\ii V(\eta^*) j_l'(kR^*)
+ \frac{\hat\Theta_2(\eta^*)}{2} \ \frac{\left(3j_l''(kR^*) + 
j_l(kR^*) \right)}{2}
+ \int_{\eta^*}^{\eta_0} d\eta 
2\dot\Phi j_l(k\Delta\eta), 
\label{alisolss}
\EN
where $R^* = \eta_0-\eta^*$ is the co-moving angular diameter
distance to the LSS.
Note that due to the presence of $e^{-\tau}$ in the 
in the last term, the range of integration is restricted
to be from about $\eta^*$ to the present.
The presence of a finite width of the LSS causes 
a contribution to the ISW effect from epochs just around last
scattering as well, usually referred to as the early ISW effect.
Once we calculate the photon brightness, 
and the baryon velocities at the epochs corresponding to
last scattering, one can calculate 
$a_l$ using \Eq{aliso} and $C_l$ from \Eq{alcl}.
Before considering the dynamics of the
the baryon-photon fluid in detail, let us first
use the \Eq{alisolss} to calculate the CMBR
anisotropies at large angular scales.

\section{Sachs-Wolfe effect and large angle anisotropies}
\label{SachsWolfe}

We wish to calculate
the anisotropies generated at large angular scales
(or small values of $l$), large enough such that the associated
spatial scales are larger than the Hubble 
radius at the LSS (i.e. $k\eta^* \ll 1$). For such scales one can 
neglect the thickness of the LSS and calculate
$a_l$ using \Eq{alisolss}. Let us also
assume that the universe is spatially flat and that
it is matter dominated by the time $\eta = \eta^*$.
The evolution of the gravitational potential $\phi$
is considered in detail in the review by Mukhanov et al. \cite{MBF}
for a variety of initial conditions, and in various epochs.
We will draw upon several of their results below.

\subsection{ Adiabatic perturbations}

Consider first adiabatic (or isentropic) perturbations, for which 
$\delta\rho_n/(\rho_n+p_n)$ is the same initially for all components. 
(Here $p_n$ is the pressure of component $n$). 
This condition is preserved
by the evolution on super Hubble scales \cite{MBF}.
As we show below, it is also preserved in the evolution of
the tightly-coupled baryon-photon fluid. 
For a flat matter dominated universe, 
the potential evolves as $\ddot\phi+(6/\eta)\phi =0$,
which implies that $\phi$ is constant in time, ignoring
the decaying mode \cite{MBF}. A detailed calculation starting from
an initial potential perturbation $\phi_i$ and following 
the evolution of the adiabatic mode from radiation era through
the matter radiation equality gives, $\phi = (9/10)\phi_i \equiv \phi_0$.
The perturbed Einstein equation also gives for the
matter density perturbation \cite{MBF},
$\delta_m= - 2\phi + (\eta^2/6)\nabla^2\phi$,
and $\vv = -(1/3)\eta\nab\phi$. For adiabatic perturbations
$\delta_R=(4/3)\delta_m$. So $\Theta_0 = \delta_R/4 =
\delta_m/3 = -(2/3)\phi +(\eta^2/18)\nabla^2\phi \to -(2/3)\phi$,
for large scales, such that $k\eta \ll 1$.
So the Fourier co-efficient $\hat\Theta_0 +\Phi = -(2/3)\Phi + \Phi
=\Phi/3$. Further since $V =-(\ii/3)(k\eta)\Phi$, this dipole term
in \Eq{alisolss} is negligible compared to the monopole term
$\hat\Theta_0 +\Phi$.  Also because of tight coupling
and negligible thickness to the LSS there is
negligible quadrupole component to $\hat\Theta_2$ for $k\eta \ll 1$. 
On super Hubble scales, for adiabatic perturbations
one then has 
\EQ
a_l = a_l^{SW} = \frac{1}{3}\Phi_0 \ j_l(kR^*); \quad
C_l^{SW} = \frac{2}{\pi} \int {dk \over k} 
\frac{k^3 \bra{\vert \Phi_0(k)\vert^2}}{9} \ j_l^2(kR^*)
\label{SW}
\EN
The above $C_l$, which describes the
CMBR anisotropies on large scales due to initially
adiabatic potential perturbations,
was first derived by Sachs and Wolfe \cite{SW67}
and is referred to as the Sachs-Wolfe effect.
For a power law spectrum of potential perturbations,
with $\Delta_\phi^2 = 
k^3\vert \Phi_0(k)\vert^2/(2\pi^2) = A_\phi^2(k_0) (k/k_0)^{n_s-1}$,
one gets
\EQ
C_l^{SW} = \frac{2A_\phi^2}{9} \left(\frac{1}{k_0R^*}\right)^{n_s-1}
\frac{2^{n_s-4}\Gamma(3-n_s)\Gamma((2l+n_s-1)/2)}
{\Gamma^2((4-n_s)/2)\Gamma((2l+5-n_s)/2)}.
\label{swgen}
\EN
(In the above equation $\Gamma(x)$ is the usual gamma function).
In theories of inflation, one obtains a nearly scale
invariant spectrum corresponding to $n_s=1$. For this
case, one gets a constant 
\EQ
\frac{l(l+1) C_l^{SW}}{2\pi} = \left(\frac{A_\phi}{3}\right)^2.
\EN
It is this constancy of $(l(l+1)C_l)/2\pi$ for scale invariant spectra that
motivates workers in the field to use this combination to present 
their results. For power law spectra, 
the recent WMAP results by themselves,
favor a nearly scale invariant spectral index with $n_s = 0.99 \pm 0.04$,
but when other large scale structure data is added slightly lower
values of $n_s$ are favored \cite{spergel03}. 
Spergel et al \cite{spergel03} also explore more complicated, 
running spectral index models, for fitting
the results from WMAP, other fine scale CMB experiments
and large scale structure data. A recent study combining
CMB and large scale structure data favors
a scale invariant spectrum, with $n_s = 0.98\pm 0.02$ with
$dn_s/d\ln k = 0.003 \pm 0.01$i \cite{seljaketal}.  Slightly different
set of parameters are derived by \cite{tegmark04},
when they combine the WMAP data with the SDSS results.

One can relate the normalization constant $A_\phi$ to the scalar normalization
$A$ used in CMBFAST and by the above authors. 
Verde et al \cite{verde03}
give $\Delta_\phi^2(k_0) = 
(800\pi^2/T) A^2$, where $T =2.725 \times 10^6 \mu \K$
and $k_0=0.002  \Mpc^{-1}$. For $n=1$, and  
adopting a value $A=0.9$, the best fit value 
for WMAP data alone, gives $A_\phi \sim 3 \times 10^{-5}$,
in agreement with the earlier COBE results. 
We can also relate $A_\phi$ to the normalization of the matter
power spectrum $\delta_H^2 = k^3 P(k)/(2\pi^2)$, evaluated at $k=H_0$.
We have $\delta_H = (2/3) A_\phi 
(D_1(a=1)/\Omega_m)^{1/2}$, where $D_1(a)$ is the growth factor 
\cite{dodelson}. For a flat matter dominated model
one would then get $\delta_H \sim 2 \times 10^{-5}$ consistent
with earlier COBE results.

\subsection{ The isocurvature mode}

The Sachs-Wolfe effect in theories which
begin initially with isocurvature perturbations can
be computed in an analogous manner. 
Suppose for example, one assumes that the universe
has two dominant components, radiation 
with density $\rho_R$ and dark matter with density $\rho_m$.
Then the total density perturbation will be
$\delta_T = (\rho_R\delta_R + \rho_m\delta_m)/(\rho_R + \rho_m)
= (\delta_R + y\delta_m)/(1+y)$. Here we have
defined $y = \rho_m/\rho_R = a(\eta)/a(\eta_{eq})$
with $\eta_{eq}$ the epoch of matter-radiation equality.
In such a model there is also an independent
'isocurvature' mode, where the initial curvature
fluctuation is zero, but there are non-zero fluctuations in the
'entropy per particle', 
$S  \propto n_R/n_m \propto T^3/\rho_m$. This entropy fluctuation
is characterized by $\sigma = \delta S/S = 3\Theta_0 -\delta_m$.
In terms of $\delta_T$ and $\sigma$, we have
$\delta_R = (\delta_T(1+y) + y\sigma)/(1 + 3y/4)$.
For such isocurvature initial conditions
on super Hubble scales, the initial value of 
$\sigma = \sigma_i$ is preserved, and this drives
the generation of curvature or potential perturbations in the
radiation dominated era (see for example \cite{MBF}). 
The resulting potential fluctuations freeze after
matter domination, on super Hubble scales, and are given by
$\phi = \sigma_i/5$ with an associated
density perturbation $\delta_T = -2\phi = -2\sigma_i/5$.
Also at matter domination, with $y\gg1$, one has
$\delta_R \to (4/3)(\delta_T+\sigma)$.
So $\Theta_0=\delta_R/4 \to (1/3)(-2\phi+5\phi) = \phi$
and $\Theta_0+\phi = 2\phi$. Then the associated
CMBR anisotropies due to the monopole term in
\Eq{alisolss} is $a_l = (\hat\Theta_0 +\Phi)j_l(kR^*)
= 2\Phi j_l(kR^*)$. The associated dipole and
quadrupole can again be neglected for super horizon
scales. For the same amplitude of potential
perturbations at the epoch of last scattering, isocurvature
initial conditions therefore lead to $6$ times larger
temperature anisotropies on large scales.
We have considered here only a two component system;
for several components, 
several independent modes of perturbations
can obtain with 
a variety of associated CMBR anisotropies \cite{mood_buc_tur}.

\subsection{ The integrated Sachs-Wolfe effect}

If the potential $\phi$ were to change with time
after decoupling, we see from \Eq{alisolss} that further
anisotropies can be generated at large angular scales.
This effect is known as the integrated Sachs Wolfe effect,
and typically arises in open universes, or in a flat universe
with dark energy/cosmological constant, wherein the
potential decays after the universe is dominated by
curvature or dark energy respectively. The gravitational
potential is also traced by other measures of large scale structure,
There has therefore been considerable interest in
checking whether there is a large angle cross-correlation between
the temperature anisotropies (some of which will arise due the ISW
effect) and other measures of large scale
structure, with some tentative detections \cite{isw_de}.

The Sachs-Wolfe and the ISW effects are dominant
on scales larger than a few degrees, or $l < 20$ or
so as schematically indicated in Figure 1. In order to understand
the smaller scale anisotropies we have to have study
in greater detail the baryon-photon dynamics, to which we now turn.

\section{ The Baryon-Photon dynamics}

We have already derived the equation for the perturbed
brightness for the photons. To complete the description of the 
baryon-photon system we have 
to also write down the continuity and Euler equations for the Baryons 
in the perturbed FRW universe. The continuity equation for the baryon 
density perturbation $\delta_B$ is 
\EQ
\dot\delta_B + \nab\cdot\vv = 3\dot\phi
\label{bcont}
\EN
where term on the RHS takes account of the variation of
the spatial volume due to the perturbed potential.
In the baryon Euler equation, we include 
the force exerted by the radiation on the Baryons due to 
$\gamma-e$ collisions.
This force is most simply calculated as the negative of the 
rate of momentum density transfer to the photons by the electrons.
The change in the momentum density of photons per unit conformal time 
to linear order in the perturbations is given by
\EQ
\frac{dT_0^{\beta}}{d\eta} =\int \frac{d^3 p_{\alpha}}{\sqrt -g}
\frac{p^{\beta}p_0}{p^0} \left(\frac{df}{d\eta}\right)_{coll} =\frac{1}{a^4} 
 \int p^3 dp d\Omega n^{\beta} \ c(f)
\label{dtmom}
\EN
where $(df/d\eta)_{coll} = c(f)$ is the change in photon momentum density
due to collisions calculated in \Sec{boltz}. The momentum transfer to the 
electrons will be negative of the value calculated in \Eq{dtmom}.  
The radiative force density exerted on the electrons (and hence the baryons) 
by the radiation is then 
\EQ
\ff_{rad} = 
-\frac{4\pi}{a^4}\int \frac{d\Omega}{4\pi}
\left[n_e\sigma_T a \frac{\rho_R a^4}{4\pi}
\left(i_0 + \frac{i_{2m}Y_{2m}}{10} +4\nn\cdot\vv -i\right)\nn
\right]
= n_e\sigma_T\rho_R a \left[\FF - \frac{4}{3}\vv\right]
\label{radf}
\EN
where $ \FF = \int (d\Omega/4\pi) \  i \ \nn$ 
is the first moment over photon directions of the fractional brightness.
So the baryons feel a force due to the radiative flux $\FF$ and 
a drag proportional to their velocity.
The Euler equation for the baryons is then 
\EQ
\rho_B [\dot\vv + {\cal H}\vv] =
-\rho_B\nabla\phi -\nabla p_B
+n_e\sigma_T\rho_R a \left [{\bf F} - {4\over 3}{\bf v}\right]
\label{EUL}
\EN
where ${\cal H} = (da/d\eta)/a$ and $p_B$ is the baryon pressure. 
The equations  \Eq{final}, \Eq{bcont}. \Eq{EUL} together with an equation of 
state for the Baryon gas form the basic set of equations for the Baryon-Photon 
system. These equations can be solved to a good approximation
in the tight coupling limit, where we consider scales
of the perturbations much larger than the photon mean-free path.
This approximation is likely to be very accurate, before the 
recombination epoch, when matter is mostly in an ionized form. 
Note that the co-moving photon mean free path $L_\gamma = (n_e\sigma_T a)^{-1}$
grows to about $\Mpc$ scales just before the decoupling epoch.
So the approximation $kL_\gamma \ll 1$ (which corresponds
to the limit $l = kR^* \ll R^*/L_\gamma \sim 10^4$), 
should hold quite accurately for most scales of interests probed 
by CMBR anisotropies. 
One can then solve for the brightness perturbation iteratively. For this 
we first rewrite equation \Eq{final} as 
\EQ
i = i_0 + {i_{2m} Y_{2m}\over 10} + 4\nn\cdot\vv    -L_\gamma 
\left[ \frac{\partial i}{\partial \eta} +\nn\cdot\nab i 
+ 8\nn\cdot\nab\phi \right]. 
\label{finit}
\EN
(The repeated $m$ index is assumed to be summed over).
We can now write down the solution 
by iteration in powers of $L_\gamma$. We get 
\EQ
i^{(0)} = i_0 + {i_{2m} Y_{2m}\over 10} + 4\nn\cdot\vv
\EN
\EQ
i^{(1)} = i^{(0)} - L_\gamma \left[
\frac{\partial i^{(0)}}{\partial \eta} +\nn\cdot\nab i^{(0)}
+ 8\nn\cdot\nab\phi \right]
\label{fone}
\EN
Here $i^{(0)}$ and $i^{(1)}$ are iterative solutions 
to \Eq{finit} giving $i$ to the zeroth and first order 
in $L_\gamma$, respectively.
(We will later consider iteration up to the second order
when deriving Silk damping).

Consider to begin with the effects of the baryon - photon 
tight coupling to the first nontrivial order 
given by $i^{(1)}$. Taking the zeroth moment of \Eq{fone}, 
that is averaging both sides of \Eq{fone} over the 
all directions of the photon momentum, we get 
\EQ
i_0 = i_0 - L_\gamma \left [ \frac{\partial i_0}{\partial \eta} 
+ \frac{4}{3} \nab\cdot\vv \right ],
\EN
where we have used the fact that 
$\int (d\Omega/4 \pi) n_i n_j = (1/3) \delta_{ij}$.
Using $\Theta_0 = i_0/4 + \phi$, we then have 
\EQ
\frac{\partial \Theta_0}{\partial \eta} + \frac{1}{3} 
\nab\cdot\vv = \dot\phi
\label{coni}
\EN
This implies that the fractional perturbation to the photon
number density $\delta n_R/n_R = 3\Theta_0$ satisfies the
same equation as $\delta_B$.
So initially adiabatic perturbations in the baryons, 
with $\delta_B = 3\Theta_0 = (3/4) \delta_R$ 
initially, maintain this relation in the radiation era.

The first moment, (that is multiplying \Eq{fone} by ${\bf n}$ and 
integrating over the directions of photon momenta) gives
\EQ
\FF = \frac{4}{3} \vv - L_\gamma \left [{\nabla i_0 \over 3} 
+ {4\over 3}\dot\vv
+ {8\over 3}\nabla \phi \right ]
\label{firmom}
\EN
The radiative force experienced by the baryons is then 
\EQ
\ff_{rad} = \frac{\rho_R}{L_\gamma} \left [\FF -{4\over 3}
\vv \right ] = -\rho_R 
\left [{\nabla i_0 \over 3}      
+ {4\over 3}\dot\vv 
+ {8\over 3}\nabla \phi \right ]
\label{radfor}
\EN
So the Euler equation for the baryons,
after substituting $\Theta_0 = i_0/4 +\phi$, becomes
\EQ
[\rho_B + {4\over 3} \rho_R]\dot\vv
+\rho_B {\cal H}\vv
= - [\rho_B + {4\over 3} \rho_R]\nab\phi
-\nab( p_B + \rho_R {4\Theta_0\over 3})
\label{EULR}
\EN
We see therefore that in the tight coupling limit, the effect of 
Thomson scattering by radiation, to the leading order, is to add 
to the baryon Euler equation : 
(i) a radiation pressure gradient term with $p_{rad} = \rho_R 4\Theta_0 /3
= \rho_R \delta_\gamma/3$, 
(ii) an extra inertia due to the radiation by
adding a mass density $(4\rho_R/3)$, 
to the inertial term in the LHS of \Eq{EULR}
and to the gravitational force term on the RHS. 
When the radiation energy density and pressure dominate 
over that of matter, the baryon photon fluid, in the 
tight coupling limit, behaves as though its mass density
is $(\rho_R+p_R)$ and its pressure $p_R = \rho_R/3$ is due to radiation.
The ratio of the inertia due to baryons and that due to radiation 
is given by $R=3\rho_B/4\rho_R \approx 0.6 (\Omega_bh^2/0.02)
(z/10^3)^{-1}$. So baryon inertia cannot be neglected. (On the other hand
the fluid pressure can be neglected compared to the radiation pressure,
since the thermal speed in the fluid is much smaller than $c/\sqrt{3}$).

On taking the time derivative of the continuity equation \Eq{coni}, 
substituting for $\dot\vv$ from the Euler
equation \Eq{EULR}, and taking its Fourier transform, we get
\EQ
\frac{\partial^2\hat\Theta_0}{\partial\eta^2}
+ \frac{\dot R}{1+R}\frac{\partial\hat\Theta_0}{\partial\eta} 
+ \frac{k^2\hat\Theta_0}{3(1+R)} = -\frac{k^2\Phi}{3} + 
\ddot\Phi +\frac{\dot R}{1+R}\dot\Phi
\label{osseq}
\EN
We see that the baryon photon fluid can undergo
acoustic oscillations, driven by the potential,
and with an effective sound speed $c_s = 1/\sqrt{3(1+R)}$.
If the baryon inertia were subdominant, with
$R\to 0$, $c_s \to 1/\sqrt{3}$ which is the
sound speed for a highly relativistic fluid.
The baryon inertia leads to a reduction of $c_s$
from this extreme relativistic value.
The oscillator equation \Eq{osseq} can also
be cast in a more suggestive form (cf.Eq. 16 of \cite{hudodr}),
\EQ
c_s^2\frac{\partial}{\partial\eta}\left(c_s^{-2}
\frac{\partial\hat\Theta_0}{\partial\eta}
\right) + c_s^2 k^2 \hat\Theta_0 = -\frac{k^2\Phi}{3}
+c_s^2\frac{d}{d\eta}\left(c_s^{-2}\dot\Phi \right).
\label{hudod}
\EN
We will use the solution of the oscillator equation \Eq{osseq}
to discuss the imprint of the acoustic waves on $C_l$. 

\section{ Acoustic peaks}

The acoustic oscillations of the baryon-photon fluid lead
to a rich structure of peaks and troughs in the CMB anisotropy
power spectrum, on sub degree angular scales (or $l > 100$).
To understand their basic features, let us look
at an approximate solution of the oscillator equation \Eq{osseq}.
To begin with let us neglect the slow variation of
$R$ with time, compared to the oscillation frequency $kc_s$.
Then we can rewrite \Eq{osseq} as 
\EQ
\frac{\partial^2(\hat\Theta_0+\Phi)}{\partial\eta^2} +
k^2c_s^2(\hat\Theta_0+\Phi) = -k^2c_s^2 R \Phi +
2\frac{\partial^2\Phi}{\partial\eta^2}
\label{ossa}
\EN
Also consider first modes which enter the Hubble radius
in the matter dominated era,
for which $(\partial^2\Phi)/\partial\eta^2 \approx 0$. Then
the solution of \Eq{ossa} is
\EQ
\hat\Theta_0+\Phi = A(k) \cos kr_s(\eta) +B(k)\sin kr_s(\eta) - R\Phi
\label{os_sol}
\EN
where $r_s(\eta) = \int_0^\eta d\eta'/\sqrt{3(1+R)}$ is called
the 'sound horizon'. Note the sine and cosine oscillations
will persist in the full solution but will have a slow
damping due to a variable $R$. The $-R\Phi$ term is
the particular solution of the inhomogeneous equation.
The effect of a non-zero $R$ (called 'baryon loading') 
is to change the sound speed $c_s$ and also shift
the zero point of the oscillations of the monopole
$(\hat\Theta_0 + \Phi)$.
One needs to specify initial conditions to fix
$A(k)$ and $B(k)$ in \Eq{os_sol}. Note that as $\eta \to 0$,
for adiabatic or curvature perturbations, we already
showed in \Sec{SachsWolfe} that $\hat\Theta_0+\Phi \to \Phi/3$.
This fixes $A(k) = (\Phi/3) (1+3R)$. Also
in the tight coupling limit, we have from \Eq{coni},
$\ii kV = -\partial\hat\Theta/\partial\eta$. Using this
relation, and noting from \Sec{SachsWolfe} 
that for adiabatic initial conditions $V \to 0$ as $k\eta \to 0$,
fixes $B(k) =0$. Imposing these initial conditions we
have for modes which enter in the matter dominated era,
\EQ
\hat\Theta_0+\Phi = \frac{\Phi}{3} (1+3R)\cos kr_s - R\Phi; \qquad
\ii V = -\frac{3}{k}\frac{\partial\hat\Theta}{\partial\eta}
= \Phi \frac{1+3R}{\sqrt{3(1+R)}} \sin kr_s
\label{os_solb}
\EN
where we have again neglected the time variation of
$\Phi$ and $R$.

\subsection{ Radiation driving}

For modes which enter the Hubble radius during the radiation
dominated era, one cannot neglect the variation in
the gravitational potential $\Phi$.  The comoving 
wavenumber $k_{eq}$, corresponding to the 
Hubble radius at matter-radiation equality, is $k_{eq} = 
H(z_{eq})/(1+z_{eq}) = (2\Omega_m H_0^2 z_{eq})^{1/2}$,
and modes with $k > k_{eq}$ enter the Hubble radius
in the radiation dominated era.
During radiation domination, for the fluid which
has an equation of state $p=\rho/3$, as would obtain
for the tightly coupled baryon-photon fluid, the
Einstein equations give $\ddot\Phi + (4/\eta)\dot\Phi
+ (k^2/3)\Phi = 0$ (dots as before denote derivatives
with respect to conformal time) \cite{MBF}. The solution for 
'adiabatic' initial condition is
\EQ
\Phi(k,\eta) = \frac{3}{(\omega\eta)^3} \left[\sin\omega\eta
-\omega\eta\cos\omega\eta\right]\Phi_i(k)
\label{phi_rad}
\EN
where $\omega = k/\sqrt{3} = kc_s$ is the frequency of
the acoustic waves and $\Phi_i(k)$ the initial
potential perturbation on super horizon scales. 
(Note that during radiation domination $c_s = 1/\sqrt{3}$).
One sees that at early times on super horizon scales,
with $k\eta \ll 1$, $\Phi \to \Phi_i$ where as once a
mode enters the Hubble radius,
the potential decays with time, going asymptotically
as $\Phi(k,\eta) \to -(3\cos\omega\eta)\Phi_i/(\omega\eta)^2$
for $k\eta \gg 1$. This decay of $\Phi$ causes extra driving of
the acoustic oscillations for such modes. We can estimate the 
effect of this extra driving by directly solving for
the associated density perturbation $\delta_R
=\delta\rho_R/\rho_R$  using the Einstein equations (cf. \cite{MBF});
$4\pi G a^2 \delta\rho_R = -k^2\Phi - 3{\cal H} \dot\Phi - 3{\cal H}^2\Phi$,
and $3{\cal H}^2 = 8\pi G a^2 \rho_R$.
For $k\eta \ll 1$ one gets
$\delta_R \to -2\Phi_i$, giving an initial value
for the monopole $(\hat\Theta_0 +\Phi) = \delta_R/4 + \Phi \to \Phi_i/2$.
On the other hand, after a mode enters the Hubble radius, 
one has asymptotically, 
$\delta_R \to -(2k^2\eta^2/3) \Phi(k,\eta) = 6\Phi_i(k) \cos (k c_s\eta)$
for $k\eta \gg 1$. So a mode which enters the Hubble
radius early in the radiation dominated era
has acoustic oscillations with 
\EQ
\hat\Theta_0 + \Phi \to \hat\Theta_0
=\delta_R/4 = \frac{3}{2} \Phi_i(k) \cos k c_s\eta .
\label{ac_rad}
\EN
The amplitude of the oscillation, 
is therefore enhanced
relative to a mode entering in the matter dominated era, 
by a factor $(3\Phi_i/2)/(\Phi_0/3) = 5$, where we have used
$\Phi_0 = (9/10)\Phi_i$.
This enhancement is referred to in the literature as
'radiation driving' \cite{husug95,husugsil}. 
The factor of $5$ derived above
gets modified to about $4$, if we include the neutrino 
component \cite{husug95}. It is also 
valid  only in the asymptotic limit of very small scales
and ignores the damping effect to be discussed below.
Further the modes which are seen as the first few peaks
in the $C_l$ spectrum have $k/k_{eq}$ larger than unity only by
a modest factor, and so the enhancement is smaller.
Nevertheless, the rise from the Sachs-Wolfe plateau of the
$C_l$ versus $l$ curve as $l$ increases from a few $10$'s to 
above $100$ or so, as displayed in Figure 1, 
is dominated by this radiation driving effect.

Note that due to the decay of the potential $\Phi$, 
the baryon loading term $R\Phi$ in \Eq{os_solb} is
absent for modes which enter the Hubble radius well into radiation
domination; so if one does see the effect of baryon loading in
the $C_l's$ at higher $l$, this would be a firm evidence
for the importance of a dark matter component in the universe
(see below).

\subsection{Silk damping}

So far we have ignored the effects of departures
from tight coupling. This departure introduces viscosity and
heat conduction effects, and associated damping of the acoustic
oscillations on small scales, worked out by Silk \cite{silk_damp}. 
To calculate Silk damping effects, one needs to go 
the second order in $L_\gamma$. 
We give a detailed derivation, starting
from the Boltzmann equation in \App{silk}.
In this derivation we neglect the anisotropy of
the Thomson scattering, and also the effects of 
$\phi$. (The scales for which damping is important,
enter in the radiation era, and so $\phi$ 
decays as explained above).

For plane wave solutions of the form,
\EQ
{\bf v} = {\bf V} \exp(\ii\kk\cdot\xx + \int^\eta \Gamma d\eta') ; \quad
\Theta_0 = \hat\Theta_0 \exp(\ii\kk\cdot\xx + \int^\eta \Gamma d\eta')
\label{forier}
\EN
we derive the dispersion relation
\EQ
\Gamma = \pm \ii k c_s
-\frac{k^2 L_\gamma}{6(1+R)^2} 
\left[R^2 + \frac{4}{5} (1+R)\right]
\label{disp}
\EN
To the first order in $L_\gamma$, the baryon-photon
acoustic waves suffer a damping, with the damping rate being 
larger for larger $k$ or smaller wavelengths. 
This damping effect \cite{silk_damp}, 
is referred to in the literature as Silk damping,   
(If one takes into account the anisotropy of the Thomson scattering
one gets $16/15$ instead of $4/5$ in the last factor above \cite{kaiser}).
Silk damping introduces an exponential damping factor
$\exp-(k/k_D)^2$ into the sine and cosine terms of \Eq{os_solb},
where the damping scale $k_D$ is determined by, 
\EQ
k_D^{-2} = \int_0^\eta d\eta'
\frac{L_\gamma}{6(1+R)^2} \left[R^2 + \frac{4}{5} (1+R)\right]
\label{silkkd}
\EN
(Also since modes with $k > k_D$, 
for which Silk damping becomes important,
enter in the radiation dominated era and their
potential $\Phi$ has already decayed significantly;
so the $R\Phi$ term for such modes is not important).
Since $R$ grows to at most $\sim 0.5$ by decoupling, the
Silk damping scale $k_D^{-1} \sim [\eta^*L_\gamma(\eta^*)]^{1/2}$
by the last scattering epoch, or the 
geometric mean of the comoving photon mean free path and the
the Hubble scale at last scattering.

\subsection{Putting it all together}

We can now put all the above ideas together to explicitly write
$C_l$ incorporating the baryon-photon oscillations. For scales
much larger than the thickness of the LSS it suffices
to use \Eq{alisolss} for $a_l$,
substituting the tight coupling expressions in \Eq{os_solb},
for $\hat\theta_0 + \Phi$ and $\ii V$. (The quadrupole term has
negligible effect in the limit of tight coupling).
The resulting $a_l$ is substituted into
\Eq{alcl} to compute $C_l$. 
Then we have for the anisotropy power spectrum,
\EQ
C_l = \frac{2}{\pi} \int \frac{dk}{k} \frac{k^3\bra{\vert\Phi_0\vert^2}}{9}
\left[\left\{ E(k) (1+3R)\cos k\eta_s^* - 3R \right\} j_l(kR^*)
+\left(\frac{E(k) \sqrt{3}(1+3R)}{1+R} \sin k\eta_s^* \right)j'_l(kR^*)
\right]^2
\label{fincl}
\EN
Note that \Eq{os_solb} only
describes accurately modes which enter in the matter dominated era.
For modes which enter the Hubble radius during radiation domination,
one has to take account of the $k$ dependent 
enhancement due to radiation driving. Also for large $k$ we have
to take account of Silk damping.
These effects can only be accurately incorporated
in a numerical solution for $C_l$. 
However many of the physical effects governing the properties of
the acoustic peaks can be illustrated 
without such a detailed solution, keeping in mind
that the co-efficients of the oscillatory terms
will have an extra $k$ dependence due to
radiation driving and Silk damping. The fudge factor $E(k)$ has
been incorporated into \Eq{fincl} to remind ourselves of the existence
of these effects.

It is also important to recall that
$j_l(kR^*)$ is a function sharply peaked at
$kR^* \sim l$. So for any given $l$, the $k$ integral 
is dominated by modes which satisfy $k \sim l/R^*$.
On the other hand, the function $j'_l(x)$ is not as strongly 
peaked as $j_l(x)$ and has also a much smaller amplitude
compared to $j_l$ (see for example \cite{husug95,huwhite97}).
So the contribution from the doppler term (which contains 
$j'_l(kR^*)$), is subdominant compared to the term
depending on the temperature and potential (which
contains $j_l(kR^*)$). We can now use \Eq{fincl} to understand 
various features in the $C_l$ spectrum. 

\begin{itemize}

\item
The CMBR power spectrum, or $C_l$ has a series of
peaks whenever the monopole term is maximum, that is
when $\cos (kr_s^*) =\pm1$, where $r_s^* = r_s(\eta^*)$
is the sound horizon at last scattering. This obtains
for $kr_s(\eta^*) = n\pi$, where $n$ is an integer; 
or for $ l \sim kR^* =  nl_A$, where we define
$l_A = \pi (R^*/r_s(\eta^*))$.
These acoustic peaks were a clear theoretical
prediction from early 70's \cite{peeb_yu,sun_zel70}; they used
to be called doppler peaks, but note that the doppler
term is subdominant compared to the temperature and potential
contribution to $C_l$. 
The peak structure for a standard $\Lambda$CDM model is shown in
Figure 1.

\item
The location of the first peak depends sensitively 
on the initial conditions, (isocurvature or adiabatic) and 
also most importantly on the curvature of the universe.
The current observations favor a flat universe.
For a flat geometry, the location of the first peak
can be used to measure the age of the universe.

\item
For isocurvature initial conditions the monopole
term would have $\sin (kr_s^*)$, which would be maximum
at $kr_s^* = (2m+1)\pi/2$, where $m=0,1,2..$. 
The peak at $kr_s^*=\pi/2$ is generally hidden.
The first prominent peak for isocurvature initial condition is at
$kr_s^*=3\pi/2$, and so occurs at larger $l$ than for adiabatic 
perturbations; present observations favor adiabatic initial
conditions.

\item
Almost independent of the initial conditions
the spacing between the peaks is $\sim l_A $.

\item
Due to a non zero baryon density, that is a non zero $R$, the
peaks are larger when $\cos (kr_s^*)$ is negative,
since in this case, the cosine term and the $-3R$ term in 
\Eq{fincl} add. Due to this effect of 'baryon loading', 
the odd peaks, with $n=1,3,..$ have larger amplitudes than the even peaks
with $n=2,4,..$. 

\item
The radiation driving effect, as we explained earlier causes the $C_l$ curve
to rise above the Sach-Wolfe signal for $l$ values corresponding 
to the acoustic peaks (cf. Figure 1). Note also that the $R\Phi$ term would
be absent, if the the scale corresponding to
a given peak enters during radiation domination, such that the
potential $\Phi$ has decayed by the epoch $\eta^*$.
Indeed the observed existence of a 3'rd peak almost
comparable in height to the 2'nd is an indication of
the importance of (dark) matter in the universe.

\item
Silk damping cuts of the $C_l$ spectrum exponentially
beyond $l \sim k_DR^* \sim 1500$ (cf. Figure 1).
There is also damping of the $C_l$ spectrum due to the finite
thickness of the last scattering surface. The scales for
both damping are similar. The decline in $C_l$ due to
both effects has been parametrized by a $\exp[-(l/l_D)^{m_D}]$ factor,
where $l_D = k_DR^*$, and $m_D \sim 1.2$ \cite{hu_white97,huetal2001}.

\end{itemize}

We also mention some of the other consequences of the varying
gravitational potential, for the $C_l$ spectrum.

\begin{itemize}

\item
Th effects of a varying gravitational potential lead
to the ISW effect as mentioned earlier.
This can operate both after
last scattering and during the period of recombination.
In an universe which is at present dominated
by dark energy, the potential associated with
sub horizon scales decay after dark energy domination.
The resulting increase in $C_l$ leads to the
upturn for $l < 10$, from the Sachs-Wolfe plateau seen in Figure 1.

\item
There is also an early ISW effect
for modes which enter the Hubble radius in the radiation
dominated era. However due to the $e^{-\tau}$ factor multiplying
$\dot\Phi$ in \Eq{aliso}, this contributes to $C_l$ only
for those modes whose potential's decay just before last scattering.
The early ISW causes an increase in $C_l$ for such modes.

\item
The early ISW effect partly fills in the rise to the
first peak and leads to a shift in the location
of the first acoustic peak to a lower $l< l_A$ \cite{hurev}.
Also for modes with $ k > k_{eq}$, 
entering the Hubble radius in the radiation era,
the decaying potential leads to a difference between the exact solution 
to \Eq{osseq} from the approximate solution given by \Eq{os_sol}.
This leads to a further shift in the location of the acoustic peaks,
to lower $l$ \cite{dodelson,huetal2001}.
Finally, $j_l^2(x)$ has a peak at slightly smaller $l$ than
$l=x$. All these effects lead to a shift of the peak
location to an $l$ value lower than $l=nl_A$, 
by $\sim 25 \%$ or so, which can only be calibrated by
numerical solution \cite{huetal2001} (see below).

\end{itemize}
Note that we can use both the location and the relative heights of
the acoustic peaks as a sensitive probe of the cosmological
parameters, an issue to which we now turn.

\subsection{The acoustic peaks and cosmological parameters}

The cosmological parameters which have been constrained
include the curvature of the universe or
the total energy density $\Omega_T$, the baryon density 
$\omega_b= \Omega_bh^2$, dark matter density $\omega_m = \Omega_mh^2$
(which is predominantly thought to be cold dark matter),
and the slope of the primordial power spectrum $n_s$.
We outline some of these ideas, following mainly
Hu {\it et al} \cite{huetal2001} and the post
WMAP analysis of Page {\it et al} \cite{pageetal2003}.

\subsubsection{ The location of the acoustic peaks}

For the flat matter dominated universe, the conformal
time $\eta \propto a^{1/2} \propto (1+z)^{-12}$. 
If we neglect the effect of baryons, $c_s=1/\sqrt{3}$ and  
$r_s^* = \eta^*/\sqrt{3}$.
Also $R^* = \eta_0-\eta^*$, and so the acoustic scale 
$l_A = \sqrt{3}\pi (\eta_0-\eta^*)/\eta^*
\approx \sqrt{3}\pi (\eta_0/\eta^*) = 172 (z^*/10^3)^{1/2}$. 
We therefore expect the
first acoustic peak around this value. It is 
however important to also take account of the 
radiation and baryon densities before decoupling.
Radiation density increases
the expansion rate and the baryon density
decreases the sound speed and so $r_s^*$ gets altered
(cf. Eq 2 in \cite{pageetal2003})
\EQ
r_s(z^*) = \frac{109.4}{\sqrt{\omega_m}} \  
\left(\frac{z^*}{10^3}\right)^{-1/2}
\frac{1}{\sqrt{R^*}}
\ln \frac {\sqrt{1+R^*} +\sqrt{R^*+r^*R^*}}{1+\sqrt{r^*R^*}} \Mpc.
\label{rs}
\EN
Here $r^* = \rho_R(\eta^*)/\rho_m(\eta^*)
\approx 0.3 (\omega_m/0.14)^{-1} (z^*/10^3)$. 
Also for a universe with non-zero curvature, 
in determining the mapping between $l$ and $k$,
its necessary to replace the
comoving angular diameter distance $d_A= R^* = \eta_0-\eta^*$
corresponding to a flat universe, by 
$d_A$ applicable to a general cosmology. This is given by,
\cite{husug95,paddycos}
\EQ
d_A \approx \frac{6000}{\sqrt{\omega_m}} \ d \Mpc ;  \qquad
d = \frac{[1+\ln(1-\Omega_V)^{0.085}]^{1+1.14(1+w)}}
{\Omega_T^{(1-\Omega_V)^{-0.76}}}.
\label{dang}
\EN
Here $\Omega_V$ is the ratio of the dark energy to the critical
energy density, and $w$ the dark energy equation of state parameter
($w=-1$ for the cosmological constant). 
For a flat $\Lambda$CDM cosmology with $\Omega_V = 0.73$ and 
$\Omega_T =1$ one gets $d \sim 0.89$.
Using $l_A=\pi d_A/r_s^*$, we see that the $\omega_m$ dependence
cancels out and 
\EQ
l_A = \frac{\pi d_A}{r_s^*} \approx
172 \ d \left(\frac{z^*}{10^3}\right)^{-1/2}
\left[\frac{1}{\sqrt{R^*}}
\ln \frac {\sqrt{1+R^*} +\sqrt{R^*+r^*R^*}}{1+\sqrt{r^*R^*}}
\right]^{-1}
\label{lafin}
\EN
Note that for a flat universe ignoring the effect
of baryons and radiation, one then gets $l_A \sim 172$,
as before. But with $\omega_b=0.02$, $\omega_m=0.14$, 
even for a flat universe, $l_A \sim 300$ and so is much larger.
We note from \Eq{lafin} that the acoustic scale is most sensitive to 
the value of $\Omega_T$, the total density parameter.

Further, the location of the first peak 
is shifted from $l_A$ because of the effects of potential decay
(as described above),
which becomes important for modes with $k > k_{eq}$, entering
the Hubble radius during radiation era. The comoving 
wavenumber $k_{eq}$ corresponds to
$l=l_{eq} = k_{eq}d_A$, where \cite{husug95}
\EQ
l_{eq} = (2\Omega_m H_0^2 z_{eq})^{1/2} d_A 
\approx 164 \ d \ \left(\frac{\omega_m}{0.14}\right)^{1/2}.
\label{leq}
\EN
One needs to work out the exact shift numerically;
For a scale invariant model, with $n_s=1$ and 
$\omega_b=0.02$, Hu {\it et al} \cite{huetal2001} give
$l_n = l_A(n-\psi)$ where $\psi \sim 0.267 (r^*/0.3)^{0.1}$, 
and for better accuracy one replaces $0.267$ with $0.24$ for $l_2$
and $0.35$ for $l_3$.
For example, for a flat $\Lambda$CDM cosmology with $\omega_b=0.02$, 
$\omega_m=0.14$, $\Omega_T=1$, $w=-1$, and taking account
of the phase shift, the first peak is predicted to be located
at $l_1 \approx 220$.
For the WMAP data, the measured value of
the $l_1 = 220.1 \pm 0.8$. 
So the data is indeed consistent with such a flat universe.
(The peaks also get affected mildly by the tilt in the power spectrum 
from $n_s=1$). 

However one should caution that $l_1$ alone does not determine the
geometry; one needs some idea of $\Omega_m$ and $\Omega_b$
which can be got from the full WMAP data.
There still remains potential degeneracies
\cite{bond_efs_teg_1997,zal_sperg_sel97,efs_bond99}, whereby the peak location
can be left unchanged by simultaneous variation in 
$\Omega_m - h$ space and $\Omega_m- \Omega_\Lambda$ space. 
If one imposes $h > 0.5$
as seems very reasonable, one infers $0.98 < \Omega_T < 1.08$
($95\%$ confidence level) \cite{spergel03}.
For the HST Key project measurement of $H_0$
as a prior, one gets $\Omega_T = 1.02 \pm 0.02$.
The observations strongly favor a flat universe.
Also from the inferred values of $\omega_b$ and $\omega_m$ from
the full data, one gets an acoustic scale $l_A \sim 300$.
If one assumes a flat universe, it turns out that
that the position of the first peak is directly
correlated with the age of the universe.
The WMAP data gives $t_0 = 13.6 \pm 2$ yr for
the $\Lambda$CDM model \cite{spergel03}

Finally, the whole $C_l$ spectrum is damped strongly
beyond the scale $l_D = k_DR^*$. Numerically, we have from Hu {\it et al}
\EQ
l_D \approx \frac{2240 \ d}
{[(1+r^*)^{1/2} - (r^*)^{1/2}]^{1/2}}
\left(\frac{z^*}{10^3}\right)^{5/4} \omega_b^{0.24} \omega_m^{-0.11}.
\label{dampl}
\EN
For the $\Lambda$CDM model with WMAP parameters, one gets $l_D \sim 1470$.
The damping scale shows a much stronger dependence on $\omega_b$ and
the redshift $z^*$ compared to $l_A$.
The small angular scale experiments like the Cosmic Background Imager (CBI)
\cite{cbiobs} do find evidence for such a damping.

\subsubsection{ Peak heights}

The heights of the different peaks, can also be used to infer
cosmological parameters. 
We define the height of the first peak as
\cite{huetal2001}, 
$H_1 = (\Delta T_{l_1}/\Delta T_{10})^2$, that giving
its amplitude relative to
the power at $l=10$. (For the WMAP data
the height of the first peak is $\Delta T_{l_1} = 
74.7 \pm 0.5 \mu \K$). 
$H_1$ increases if (a) $\omega_m$ decreases
(because radiation driving is more effective a lower 
matter density), (b) if $\omega_b$ increases (due to the baryon
loading) (c) if one has a lower $\Omega_\Lambda$ or
higher $\Omega_T$ (because then the integrated Sachs Wolfe effect
is smaller which decreases $\Delta T_{10}$). 
Further $H_1$ can decrease if one has re-ionization
(since a fraction $\tau$ of photons are re-scattered), or
if one has a contribution from tensor fluctuations
(tensors will contribute to Sachs-Wolfe effect but not to
acoustic oscillations). Since $H_1$ depends on several effects,
there is no simple fitting formula; around $\Lambda$CDM \cite{huetal2001}
have given a crude formula for its variation with various parameters.

The height of the second peak is defined relative to the
first, as $H_2 = (\Delta T_{l_2}/\Delta T_{l_1})^2$.
This ratio is insensitive to reionization or to the overall
amplitude of the power spectrum since these
scale both peaks by the same amount. The dependence on
$\omega_m$ is also weak because radiation driving roughly
affects both peaks similarly. $H_2$ is most sensitive to
the baryon density $\omega_b$, since baryon loading increases
the first peak relative to the second. It is also sensitive
to any tilt in the spectrum, away from $n_s = 1$. From
fitting to a grid of spectra using CMBFAST \cite{cmbfast},
one has \cite{pageetal2003}
\EQ
H_2 = 0.0264 \omega_b^{-0.762} (2.42)^{n_s -1}
\times \exp [-0.476\ln(25.5\omega_b + 1.84\omega_m)^2]
\label{h2}
\EN
\EQ
\frac{\Delta H_2}{H_2} = 0.88 \Delta n_s - 0.67 \frac{\Delta\omega_b}
{\omega_b} + 0.039 \frac{\Delta\omega_m}{\omega_m}
\label{h2dep}
\EN
For the WMAP data, $H_2 = 0.426 \pm 0.015$. For a fixed
$\omega_m$ the first two terms of \Eq{h2dep} quantifies
the degeneracy in the $\omega_b-n_s$ plane.

The height of the third peak increases as $\omega_b$ increases
(baryon loading). The ratio $H_3 = (\Delta T_{l_3}/\Delta T_{l_1})^2$
is most sensitive to $n_s$ or 
any departures from scale invariance, because of the
long $l$ baseline. Hu {et al} \cite{huetal2001} give
\EQ
H_3 = \frac{2.17 \omega_m^{0.59} (3.6)^{n_s-1}}
{[1 + (\omega_b/0.044)^2][1 + 1.63(1-\omega_b/0.071)\omega_m]}
\label{h3}
\EN
\EQ
\frac{\Delta H_3}{H_3} = 1.28 \Delta n_s - 0.39 \frac{\Delta\omega_b}
{\omega_b} + 0.46 \frac{\Delta\omega_m}{\omega_m}
\label{h3dep}
\EN
These dependencies are accurate to few percent levels
for variation around the WMAP inferred parameters \cite{pageetal2003}.
WMAP does not yet clearly measure the third peak,
but from previous compilations \cite{wang03},
Page {\it et al} estimate $H_3 = 0.42 \pm 0.08$.
Note that if $n_s$ is fixed, $\omega_b$ is well constrained
by $H_2$ and then $\omega_m$ from $H_3$. For more
details we refer the reader to \cite{pageetal2003}.
We show in Figure 2 a set of $C_l$ versus $l$ curves, generated using
CMBFAST, which illustrate the sensitivity of the CMBR to
the cosmological parameters discussed above.

\subsection{ Other sources of CMB anisotropies}

So far we have concentrated on the primary temperature
anisotropies generated at the LSS; a number of effects can
generate additional anisotropies after recombination,
generally referred to as 'secondary anisotropies'.
We do not discuss these in detail;
for an extensive review see \cite{hudodr}.
Of the gravitational secondaries, we have already discussed the 
ISW effect arising from the changing gravitational potential. 
This effect is also important
if there are tensor metric perturbations, say due to
stochastic gravitational waves generated during inflation \cite{tensorsv}.
Another important gravitational secondary arises due to gravitational lensing
(cf. \cite{hudodr} and references therein). 
Scattering effects due to free electrons
along the line of sight can also produce a number of effects. 
The electrons can arise in collapsed objects like clusters
or due to re-ionization of the universe.
We discuss the effects of re-ionization later below. 
The scattering of the CMB due to the ionized electrons in
clusters of galaxies was first discussed by
Sunyaev and Zeldovich (SZ) \cite{sun_zel72}. 
The SZ effect generates power
below the damping tail in the $C_l$ spectrum, at a level which
depends on the normalization of the density power spectrum,
$\sigma_8$. (Here $\sigma_8$ is the RMS density contrast when the 
density field is smoothed over a 'top hat' sphere of radius $8h^{-1} \Mpc$). 
Recently a significant excess power was detected by the CBI 
experiment \cite{cbiobs}, 
at small angular scales ( $l > 2000$) at a level of $\sim 355 (\mu \K)^2 $.
This can arise from the SZ effect, but requires a somewhat large
$\sigma_8 \sim 1$ (cf. \cite{cbisz}), larger
than values previously assumed. 
Alternatively the CBI result may point
to new physics; it has been suggested for example that
primordial magnetic fields can be a significant
contributor to the power at large $l$ \cite{ksjdb_bc,trsksjdb}. 
Primordial tangled magnetic fields generate vortical ( Alfv\'en
wave mode) perturbations which lead to temperature
anisotropies due to the doppler effect. They also
survive Silk damping on much smaller scales
compared to the scalar modes \cite{jedam,ksjdba}.
The test of whether the CBI excess is indeed produced
by the SZ effect, will come from the spectral dependence of the
excess power (if it is due to the SZ effect, there should
be such a spectral dependence), and measurements of
polarization on these small angular scales (see below).
There are several other interesting gravitational and scattering 
secondaries which can generate temperature anisotropies,
and we refer the interested reader to the excellent review \cite{hudodr}.

\section{ Polarization of the CMBR}

\subsection{ The origin of CMB polarization}
 
It was realized quite soon after the discovery of the
CMB that it can get polarized \cite{rees68}.
Polarization of the CMBR arises due to Thomson scattering
of the photons and the electrons, basically because 
the Thomson cross section is polarization dependent.
We used in earlier sections the cross section relevant
for unpolarized light, ignoring the small effects of polarization
on the temperature evolution. Scattering of
radiation which is isotropic or even one which has a dipole
asymmetry is however not capable of producing polarization.
The incoming radiation needs to have a quadrupole anisotropy.
The general features of CMBR polarization are discussed 
in detail in some excellent reviews \cite{kosowsky,huwhite_new}.
Note that in the tight coupling limit, the radiation field
is isotropic in the fluid rest frame, and can have at most
a dipole anisotropy in the frame in which the fluid moves. The quadrupole
anisotropy is zero. However to the next order,
departures from tight coupling, 
due to a finite photon mean free path, in the presence
of velocity gradients, can generate a small quadrupole anisotropy.

A qualitative argument is as follows \cite{zal03}: 
The last scattering electron (say $O$ at $\xx_0$) sees radiation
from the 'last but one scattering' electron ($P$), roughly a 
photon mean free path ($L_\gamma$) away, say
at a location $\xx = \xx_0 +L_\gamma \nn$. Here $\nn$
is the direction from $O$ to $P$.
The velocity of the baryon-photon fluid at
$P$ is $v_i(\xx) \approx v_i(\xx_0) +L_\gamma n_j\partial_jv_i(\xx_0)$.
Due to the Doppler shift, the temperature seen by $O$ is
$\delta T(\xx_0,\nn)/T \sim n_i[v_i(\xx) - v_i(\xx_0)]
=L_\gamma n_in_j\partial_jv_i(\xx_0)$.
This is quadratic in $\nn$ and so corresponds to a quadrupole
anisotropy as seen by the last scattering electron.
The Thomson scattering of this quadrupole anisotropy 
can lead to polarization of the CMBR. The fractional polarization
anisotropy generated is $\Delta_P \sim kL_\gamma V$.

One complication is that $L_\gamma$ grows rapidly
as photons and baryons decouple during recombination.
An approximate estimate of its effect, would be to weigh the
polarization amplitude derived above, with the probability
of last scattering at a given epoch described by the visibility
function. Note that the visibility function goes as $\dot\tau e^{-\tau}$,
where  $\dot\tau = 1/L_\gamma$. So during the tight coupling
evolution, the $L_\gamma$ factor cancels
out and after the weighting one gets instead
$\Delta_P \sim k\delta\eta^* V$,
where $\delta\eta^*$ is the width of the LSS. So the effective
photon mean free path generating quadrupole anisotropy and
hence polarization of the CMB becomes $\delta\eta^*$, the average
distance photons travel between their last and last but one scattering,
during decoupling. Such an estimate is verified in a more careful 
calculation \cite{zalhar95}.

\subsection{ Describing CMBR polarization}

There is another complication that has to be handled when
dealing with polarization, the fact that polarization is not
a scalar quantity. It is conventional to describe polarization 
in terms of the Stokes parameters, $I$, $Q$, $U$ and $V$,
where $I$ is the total intensity, whose perturbed version was 
called $i$ above, and discussed extensively in earlier sections.
For a quasi-monochromatic wave, propagating in the $z$-direction,
we can describe the electric field at any point in space
as $E_x = a_x(t) \cos[\omega_0t -\theta_x(t)]$ and 
$E_y = a_y(t) \cos[\omega_0t -\theta_y(t)]$, where
the amplitudes $a_x$, $a_y$ and the phases $\theta_x$, $\theta_y$
vary slowly in time, compared to $\omega_0^{-1}$. The stokes parameters are
defined as the time averages: $I =\bra{a_x^2} + \bra{a_y^2}$,
$Q= \bra{a_x^2} - \bra{a_y^2}$, $U = \bra{2a_xa_y \cos(\theta_x-\theta_y)}$,
$V=\bra{2a_xa_y \sin(\theta_x-\theta_y)}$. Unpolarized light has $Q=U=V=0$.
The parameters $Q$ and $U$ describe linear polarization,
while $V$ describes circular polarization.
At the zeroth order the CMB is unpolarized and
its small polarization is expected to arise as explained
above due to Thomson scattering. This does
not produce circular polarization and so
one can set $V=0$.

Note that under a rotation of the $x$ and $y$ axis through
an angle $\psi$, the parameters $I$ and $V$ are invariant
but $(Q \pm \ii U)' = e^{\mp 2\ii\psi} (Q \pm \ii U)$.
So $Q \pm \ii U$ transform as a spin 2 Tensor under rotation
of the co-ordinate axis. The standard spherical harmonics
do not provide the appropriate basis for its Fourier expansion
on the sky. One then adopts
the following approach to this problem \cite{zal_sel97,KKS97}; 
construct scalars
under rotation from $Q \pm \ii U$ by using spin-lowering
($\partial^-$) and spin-raising ($\partial^+$) operators, and 
then make a standard $Y_{lm}$ expansion.
Or alternatively construct tensor ('spin' weighted) spherical harmonics,
$_{\pm 2}Y_{lm}$ by operating on the $Y_{lm}$'s twice with spin-raising
or lowering operators, and then expand
\EQ
(Q \pm \ii U)(\nn) = \sum_{lm} a_{\pm 2,lm} \  (_{\pm 2}Y_{lm})
\EN
in this basis. Alternatively $a_{\pm 2,lm}$ can also be thought
of as the $Y_{lm}$ expansion co-efficients of the spin zero
quantities, $(\partial^-)^2(Q + \ii U)$ 
and $(\partial^+)^2(Q - \ii U)$ respectively, apart from an $l$ dependent
normalization factor. The explicit expressions for the
raising and lowering operators, the spin weighted harmonics,
and the expansions in terms of these are given
in \cite{zal_sel97}. For example we can write,
\EQ
a_{\pm 2,lm} = \int d\Omega \ (_{\pm 2}Y^*_{lm}(\nn)) \ (Q \pm \ii U)(\nn)
= \left[\frac{(l+2)!}{(l-2)!}\right]^{-1/2}
\int d\Omega \ Y^*_{lm}(\nn) \left[(\partial^{\mp})^2(Q \pm \ii U)(\nn)
\right] 
\label{a2lm}
\EN
Since $a_{\pm 2,lm}$ are
expansion co-efficients of scalar quantities under rotation, they can be used 
to characterize the polarization on the sky in an 'invariant' manner.
More convenient is to use the linear combinations,
$a_{E,lm} = -(a_{2,lm} + a_{-2,lm})/2$ and 
$a_{B,lm} = \ii(a_{2,lm} - a_{-2,lm})/2$ \cite{zal_sel97,new_pen66},
and the associated real space polarization fields;
$E(\nn) = \sum_{lm} a_{E,lm} Y_{lm}(\nn)$ and 
$B(\nn) = \sum_{lm} a_{B,lm} Y_{lm}(\nn)$. The $E$ and $B$ fields
specify the polarization field ($Q$ and $U$) completely,
are invariant under rotation (just like the temperature $\Theta(n)$)
and have definite parity. Under a parity transformation,
$E$ remains invariant while $B$ changes sign \cite{new_pen66}.
The convenience of the $E$-$B$ split comes from the fact that
scalar perturbations do not produce any $B$ type polarization.
An alternative way of thinking about the $E$ and $B$ split 
is that they are the gradient and curl type components 
of the polarization tensor \cite{KKS97}. 
More details of these fascinating but somewhat
complicated ideas can be got from the two seminal papers on the
subject \cite{KKS97,zal_sel97}.

In order to describe the statistics of CMBR anisotropies fully, including
its polarization, we have to now consider not only $C_l$ due
to the temperature anisotropy $\Theta$, but also corresponding 
power spectra of $E$, $B$ and the cross correlation between
$\theta$ and $E$. Note that the cross correlation between $B$ and $\Theta$,
and $B$ and $E$ vanish if there are no parity violating effects.
Since $E$ and $B$ are rotationally invariant quantities, 
we can define the power spectra $C_l^{E}$, $C_l^{B}$ and $C_l^{TE}$ 
in an analogous way to the temperature power spectrum.
We now turn to their computation.

\subsection{ Computing the polarization power spectrum}

We focus on scalar perturbations. In this case for any given Fourier mode
$\kk$, one can define a co-ordinate system with $\kk \parallel \hat{\bf z}$,
and for each plane wave, treat the Thomson scattering as the
radiative transport through a plane parallel medium. It turns
out that only Stokes Q is generated in this frame because of azimuthal
symmetry, and its amplitude depends only on $\mu = \nn\cdot\kkk$.
The stokes parameter $U=0$ in this frame, for each $\kk$ mode.
Because $U=0$ and $Q$ is only a function of $\mu$, one has
$(\partial^-)^2(Q + \ii U) = (\partial^+)^2(Q - \ii U)$.
(From the explicit form of the spin-raising/lowering
operators give in \cite{zal_sel97}, it can be checked that
$(\partial^-)^2(f(\mu)) = (\partial^+)^2(f(\mu))$ for any 
azimuthally symmetric function which depends only on
$\mu$. Second since $U=0$, we have $(Q + \ii U)=(Q - \ii U)$). 
Therefore $a_{2,lm} = a_{-2,lm}$, and we have $a_{B,lm} =0$
for such scalar perturbations.

The Boltzmann equation including polarization is given  
by a number of authors (see for example \cite{bond_efs84,ma_bert,hu_white97}).
We will simply quote the result, got using the detailed treatment by
\cite{hu_white97}. We have
\EQ
C_l^E = \frac{2}{\pi} \int {dk \over k} k^3 \bra{\vert a_l^E(k,\eta_0)\vert^2}
; \qquad
a_l^E(k,\eta_0) = -\int d\eta \ g(\eta_0,\eta) \frac{kL_\gamma V}{3} 
\left[\frac{(l+2)!}{(l-2)!}\right]^{1/2} 
\frac{j_l(k\Delta\eta)}{(k\Delta\eta)^2}
\label{ael}
\EN
where we have expressed the quadrupole source for the polarization anisotropy,
$P = (\Theta_2 - \sqrt{6} E_2)/10$, by its tight coupling limit
$P= 2kL_\gamma V/9$ (see \cite{hu_white97}).
As argued on qualitative grounds above, polarization is sourced
by the velocity differences of the fluid, over a photon mean free path
(i.e. $kV L_\gamma$). Once again the spherical Bessel function 
$j_l(k\Delta\eta)$ in \Eq{ael} will at a given $l$, pick out
(on $k$ integration) a scale $k \sim l/\Delta\eta$
at around last scattering, while the visibility function $g$
weighs the contribution at any time $\eta$ by the probability
of last scattering from that epoch. 

Suppose we wish to estimate the polarization anisotropy
on physical scales much bigger than the thickness
of the last scattering surface, or $l \sim k R^* < 1000$ or so. 
As we explained earlier,
the visibility function goes as $\dot\tau e^{-\tau}$ whereas
the polarization source is $kL_\gamma V/3 = (kV/3)(\dot\tau)^{-1}$, 
and so in their product, $\dot\tau$ cancels and
only $e^{-\tau} (kV/3)$ would survive.
The integral over $\eta$ in \Eq{ael}, would be nonzero only
for a range of epochs of order the width $\delta\eta^*$ of the LSS.
(Note that just after recombination, the tight coupling
expression cannot be used; however there is also no
polarization for $\eta > \eta^*$ because there is negligible further
Thomson scattering).
So one expects a contribution of order $kV\delta\eta^*/3$ in doing
this integral, apart from an evaluation of the other terms
at $\eta^*$. A more rigorous analysis,
following the time evolution of the polarization
source term, gives a further factor of $\sim 1/2$ reduction,
if $\delta\eta^*$, is defined as the Gaussian width of the
visibility function \cite{zalhar95}.
Making such an approximation, and putting in the tight coupling
expression for the velocity of the photon-baryon fluid, gives
\EQ
a_l^E(k,\eta_0) = -\frac{k\delta\eta^*}{6}
\ \Phi_0 \ E(k) (1+3R)c_s \sin k\eta_s^*  
\left(\frac{l}{kR^*}\right)^2 \ j_l(kR^*)
\label{finale}
\EN
Note that again the $k$ integral to find $C_l^E$ will pick out values
of $kR^* \sim l$. We can infer a number of features of
the polarization from the above:

\begin{itemize}

\item
The magnitude of the polarization anisotropy, is of order
$\Delta P \sim 0.6 (k\delta\eta^*) (\Phi_0/3) 
= 0.6 \  l(\delta\eta^*/R^*)(\Phi_0/3)$,
where we have taken $R \sim 0.6$. Adopting $\delta\eta^* \sim 10 h^{-1} \Mpc$
and $R^* \sim 10^4 h^{-1} \Mpc$, we get 
at $l\sim 100$, a polarization anisotropy about $6\%$ of the Sachs-Wolfe
contribution (or about $2 \mu \K$). The amplitude rises with $l$,
but at large $l > l_D$ the Silk damping cuts off the
baryon-photon velocity, and so the polarization gets cut off
as $e^{-(l/l_D)^m}$, say. This gives a maximum contribution
at $l < l_D$ depending on the $m$, with peak amplitude 
of order $10\%$ of the peak temperature anisotropies.
These order of magnitude estimates
are borne out by the more detailed numerical
integration using CMBFAST shown in Figure 1.

\item
The acoustic oscillations of the baryon-photon fluid velocity
imprints such oscillations also on the polarization. The polarization
will display peaks when $\sin(kr_s^*) = \pm 1$, or for $kr_s^* = (2n+1)
(\pi/2)$, with $n=0,1,..$, 
corresponding to $l \sim (2n+1) l_A/2$. These peaks are 
out of phase with the temperature acoustic peaks, as they arise due to
the velocity, and they are sharper
(since for temperature there is a partial filling in of the troughs 
by the velocity contribution). 

\item
Both the polarization and the temperature depend 
on the potential $\Phi$, and so one expects a significant cross correlation
power $C_l^{TE}$. Further, the $j_l$ term does not significantly correlate
with $j_l'$ term in the $k$-integral for $C_l^{TE}$. So 
this cross correlation will be dominated by the product
of the temperature monopole with a $\cos(kr_s^*)$ dependence
and the polarization (of $E$ type) with a $\sin(kr_s^*)$ dependence. 
The peaks of $C_l^{TE}$ will then occur
when $\sin(kr_s^*)\cos(kr_s^*) \propto \sin(2kr_s^*) = \pm 1$,
or when $kr_s^* = (2n+1) (\pi/4)$, with $n=0,1,..$,
corresponding to $l \sim (2n+1) l_A/4$. So $C_l^{TE}$
has oscillations at twice the frequency compared to
the temperature or polarization. 
There will be shifts in the exact location of
the $C_l^E$ and $C_l^{TE}$ peaks, as for the temperature.

\item
The $E$ type polarization has been detected at a $5\sigma$ level
by the Degree Angular scale Interferometer (DASI) at $l$ values
of a few hundred, and at a level consistent with the
expectations from the detected temperature anisotropy \cite{kovac02}. 
The CBI experiment has also detected the $E$ type polarization,
with the peaks in the polarization spectrum showing the expected
phase shifts compared to the peaks of the temperature spectrum \cite{read2}.
The $TE$ cross correlation was detected at $95\%$ significance by DASI,
but there is no evidence of any $B$ type polarization.
The cross correlation has also been detected by WMAP.
The WMAP experiment has released results on 
$C_l^{TE}$, although not on $C_l^{E}$. WMAP
detects significant negative $C_l^{TE}$, at $l \sim 150$
and a positive 'peak' at $l\sim 300$. The existence
of such an anti-correlation between temperature and polarization
is an indication that there exist 'super-Hubble' scale fluctuations
on the LSS. This is interpreted as strong evidence for inflation type
models, since models which involve seeds (like cosmic strings)
can produce super Hubble scale temperature fluctuations 
(due the ISW type effects) but not the observed anti-correlation
in $C_l^{TE}$.

\end{itemize}

\subsection{ $B$ type polarization}

So far we have emphasized the $E$ type polarization, as 
scalar modes do not produce the $B$ type signal. 
However models of inflation which are thought to generate
the scalar perturbations, can also generate a stochastic background 
of gravitational waves. These tensor perturbations
and the CMBR anisotropy that they generate has
also been studied in detail \cite{tensors}, although we
will not do so here. Their effects are best separated from
the scalar mode signals, by the fact that Tensors also
lead to $B$ type polarization anisotropy \cite{zal_sel97,KKS97}. 
The temperature contribution from tensors is flat roughly upto
$l\sim 100$ after which it rapidly falls off. The polarization
contribution, produced at recombination, peaks at $l \sim 100$. 
The peak amplitude of the signal is however expected to be
quite small in general with 
$(l(l+1)C_l^B/2\pi)^{1/2} \sim 0.1\mu \K (E_{inf}/2 \times 
10^{16} {\rm GeV})^2$, where $E_{inf}$ is the energy scale of inflation.
\cite{zal03}. (An $E_{inf} \sim 2 \times 10^{16} {\rm GeV}$ 
corresponds to the ratio of the $l=2$ contribution due to tensors  
compared to scalar, $T/S \approx 0.1$).
One of the prime motivations for measuring polarization with great
sensitivity is to try and detect the contribution
from stochastic gravitational waves.
The $B$ type anisotropy can also arise due to gravitational lensing
of the CMB, even if one had only $E$ type polarization arising
from the recombination epoch \cite{zal_sel98b}.
This could set the ultimate limitation for detecting the
$B$ mode from gravity waves.
Another interesting source for $B$ type polarization are vector modes,
arising perhaps due to tangled magnetic fields
generated in the early universe \cite{trsksjdb,seshksjdb,mack,lewis}, or even
present in the initial conditions \cite{lewisb}.
Indeed if there were helical primordial magnetic fields,
at the LSS, parity invariance can be broken
and one can even generate $T-B$ cross correlations \cite{dur_tan}.

\subsection{ Reionization and CMBR polarization}

One of the surprises in the WMAP results was the detection of
a significant excess cross-correlation power $C_l^{TE}$ at low $l$, over and
above that expected if polarization was only generated at recombination.
This can be interpreted as due to the effects of re-ionization,
But one seems to need a significantly higher optical depth
to the re-ionized electrons $\tau_{ri} \sim 0.17$, and a correspondingly
high redshift for reionization $z_{ri} \sim 17$. The probes
and models of the high redshift intergalactic medium, including
the use of the CMBR as a probe of re-ionization is discussed
more fully elsewhere in this volume  by Sethi \cite{sethi04}.
We make a few qualitative remarks. 

First, note that if photons are
re scattered, due to electrons produced in re-ionization, the visibility
function will have 2 peaks; one narrow peak around
recombination, and a broader peak around the re-ionization epoch
(cf. Figure 2 in \cite{sethi04}), which depends on the
exact re-ionization history. The probability for last
scattering around the usual LSS will diminish by a multiplicative factor
$e^{-\tau_{ri}}$, where $\tau_{ri}$ is the optical depth
for electron scattering to the re-ionization epoch.
At the same time new temperature and polarization anisotropies get generated.
The most important effect is that Thomson
scattering by electrons generated during re-ionization, produces
additional polarization. Note that the quadrupole anisotropy 
at the the re-ionization epoch is likely to be much larger than 
at recombination, simply because
the monopole can free stream to generate a significant quadrupole at
the new LSS. At re-ionization redshifts
close enough to the observer, the relevant monopole becomes
the Sachs-Wolfe value $\Theta_0+\Phi = \Phi_0/3$. The quadrupole
at the re-ionized epoch $\eta_{ri}$ can then be simply estimated
by replacing $\eta_0$ in \Eq{alisolss}, by $\eta_{ri}$.
One gets $\Theta_2(\kk,\eta_{ri}) = a_2(\kk,\eta_{ri})
= (\Phi_0/3) j_2(k(\eta_{ri} -\eta^*))$.
Note that this does not have the $kL_\gamma$ suppression
factor, which obtains around recombination. 
Also $E_2$ in the polarization source term $P$ above
is negligible. In evaluating
the $E$ type polarization arising from the re-ionization,
one can substitute the resulting $P=\Theta_2/10$
in \Eq{ael} instead of $P= 2kL_\gamma V/9$; for the range
of $\eta$ where scattering by electrons generated due to
re-ionization is important.  

The resulting re-ionization contribution can be best calculated numerically,
for example using CMBFAST, as illustrated by Sethi (this volume).
But the scale where the peak in the power spectra can be estimated
noting that $a_l^E$ will involve the product $j_2(k(\eta_{ri} -\eta^*))
j_l(k(\eta_0 -\eta_{ri}))$, which contributes to the
$k$-integral dominantly when both $k(\eta_{ri} -\eta^*) \sim 2$ and
$k(\eta_0 -\eta_{ri}) \sim l$. This implies that the
re-ionization contribution to $E$ type polarization peaks at 
$l \sim 2(\eta_0 -\eta_{ri})/ (\eta_{ri} -\eta^*) \sim 10$
for the parameters appropriate for a $\Lambda$CDM cosmology and
a $z_{ri} \sim 20$.
This scale basically reflects the angle subtended by the 
Hubble radius at re-ionization. One has to also take account of the damping
due to the large width of the LSS at re-ionization, which
will shift the peak to smaller $l$. 

The $k$ integral
which determines $C_l^{TE}$, involves the product 
$j_2(k(\eta_{ri} -\eta^*)) j_l(k(\eta_0 -\eta_{ri}))j_l(kR^*)$,
the last $j_l(kR^*)$ coming from the temperature contribution
from the usual LSS. Note that $\eta_0$ is much bigger than
both $\eta_{ri}$ and $\eta^*$, and the two $j_l$ factors
will re-inforce each other for small $l$.
The cross correlation peak will occur at an $l$ similar
to the peak in $C_l^E$.
The indication from the WMAP data for significant
optical depth from re-ionization is not easy
to explain (cf. \cite{reiond}). If the preliminary
WMAP result continues to firm up with subsequent
years data, it will set very strong constraints on
the star and active galaxy formation at high redshift.
It may be also worth exploring new physical
alternatives. For example Ref~\cite{sethi_sub}, explores
the possibility that tangled magnetic fields generated in the early 
universe could form subgalatic objects at high enough redshifts
to impact significantly on re-ionization. Note that if there
is significant optical depth to re-ionization, then 
inhomogeneities at the new LSS can lead to
new secondary sources of both temperature \cite{ostriker_vish} and
polarization anisotropies \cite{polin}.
Eventually the detailed measurement of the polarization signals,
could be a very effective probe of the reionization history
of the universe \cite{kaplin}.

\section{Concluding remarks}
 
In this review we have tried to emphasize
the physics behind the generation of CMBR anisotropies.
We have given details of the computation of the primary
temperature anisotropies, and also indicated the relevant issues
for polarization. Our aim is more to introduce the budding
cosmologist to the well known (and reviewed) techniques
used to calculate the CMB anisotropies, rather than provide
an extensive survey of observations and results.
Of course, it is the existence of very good observational
data that makes the effort worthwhile. 
Clearly the CMB is and will continue
to be a major tool to probe structure formation and
cosmology. We have already learned a great deal from the
detailed observations of the degree and sub degree scale 
temperature anisotropies, particularly the acoustic peaks.
The exploration of small angular scale anisotropies is just at
a beginning stage and holds the promise of revealing
a wealth of information, on the gastrophysics of structure formation.
The future lies in also studying in detail the polarization of
the CMBR. Already WMAP results have revealed a surprisingly
large redshift for the re-ionization of the universe. Polarization
will also be a crucial probe of the presence of gravitational waves.
We can expect in the years to come much more
information on cosmology from WMAP, future missions like PLANK 
and other CMB experiments, with the possibility of more surprises!

\begin{acknowledgements}

I thank John Barrow and T. R. Seshadri for many discussions 
and enjoyable collaborations on the CMB over the years. 
I also thank T. Padmanabhan for encouragement, making me give 
various talks on the CMB over the years, and extracting this review!

\end{acknowledgements}

\appendix
\section{The collision term: details}
\label{collision}

Consider the integral over the collision term on the RHS of
\Eq{btli}. We have $ \int p^3 dp \bar c(f) = n_e \sigma_T( A_1 + A_2) $
where
\EQ
A_1 = \int p^3dp { d\Omega^{\prime}\over 4\pi} 
[\bar f(\bar p,\bar{\bf n}^{\prime})- \bar f(\bar p,\bar{\bf n})];
\quad 
A_2 = \int p^3dp { d\Omega^{\prime}\over 4\pi}
 {1\over 2}P_2(\bar{\bf n}\cdot\bar{\bf n}^{\prime})
 [\bar f(\bar p,\bar{\bf n}^{\prime})- \bar f(\bar p,\bar{\bf n})]
\label{aonetwo}
\EN
are respectively, the isotropic and anisotropic contribution 
to the collision term. 
From the invariance of the scalar $p_iu^i$, where $u^i$ is the four 
velocity corresponding to the bulk motion 
of the electron (Baryonic) fluid we can show that
\EQ
p = a (1 + {\bf n}.{\bf v})(1 - \phi) \bar p 
\label{dopp}
\EN
(We have used here the fact that in the fluid rest frame the components of 
$\bar u^i = (1,0,0,0)$, while $u^i = (\gamma_v/\sqrt{g_{00}}
, \gamma_v \vv/\sqrt{\vert g_{\beta,\beta}\vert})$ with
$\gamma_v = (1 - v^2/c^2)^{-1/2}$).
We split $A_1= I_1 - I_2$ with
\EQ
I_1 = \int p^3dp { d\Omega^{\prime}\over 4\pi} 
\bar f(\bar p,\bar{\bf n}^{\prime})
\quad
I_2 = \int p^3dp { d\Omega^{\prime}\over 4\pi} \bar f(\bar p,\bar{\bf n})] 
\label{ionetwo}
\EN

For evaluating $I_2$ we use the fact that $f$ is a scalar, that is
$\bar f(\bar p, \bar{\bf n}) = f(p, {\bf n})$. Also the 
integrand of $I_2$ does not depend on $\bar{\bf n^{\prime}}$. So we have
$I_2 = \int p^3 dp f(p,{\bf n}) = (\rho_R a^4/ 4\pi)[1 + i]$.
For evaluating $I_1$ we stay in the initial electron rest frame and 
transform the integral over $p$ to one over $\bar p$ using \Eq{dopp}.
We get
$I_1= (a^4/4\pi) (1-4\phi)(1 + 4{\bf n}\cdot{\bf v})\bar \rho$
where we have used the fact that
$\bar\rho = \int d^3 \bar{\bf p}\bar p \bar f(\bar{\bf p})$
is the energy density of radiation in the fluid rest frame.
Using the invariance of $T^i_ku_iu^k$, 
and from the fact that the components of
both $u^i$ and $T^i_k$ which involve one spatial index 
are of order $v/c$, 
we can check that $\bar \rho = \rho + O(v^2/c^2)$.
Since, $\rho = \rho_R(1+4\phi)(1+i_0)$, to linear order
$I_1= (a^4\rho_R/4\pi) (1 + 4{\bf n}\cdot{\bf v} + i_0)$.
So
\EQ
A_1 = {a^4 \over 4\pi}\rho_R
\left[ i_0 + 4\nn\cdot\vv - i \right] 
\label{ifrt}
\EN
To simplify $A_2$, we use the addition theorem 
for spherical harmonics to write 
\EQ
A_2= \sum_{m=-2}^2 \frac{Y_{2m}(\bar\nn)}{10}
\int p^3dp \  d\Omega^{\prime} 
\ Y_{2m}^*(\bar\nn') [\bar f(\bar p,\bar\nn')-\bar f(\bar p,\bar\nn)] 
={a^4\rho_R \over 4\pi}\frac{1}{10}\sum_m [Y_{2m}(\nn) i_{2m}],
\label{atwo}
\EN
where $i_{2m} = \int d\Omega Y^*_{2m} i(\xx,\eta,\nn)$. 
In evaluating $A_2$ we have used the fact that
the term $\bar f(\bar p,\bar\nn)$ does not contribute
to the integral over $d\Omega^{\prime}$. Also
writing $\bar f(\bar p,\bar\nn') = f(p,\nn') 
= f_b(p) + f_1( p,\nn')$, the
$f_b$ term gives zero contribution.
And since $f_1$ is already first order in perturbations,
we can evaluate $A_2$ by replacing  $\bar \nn$ and $\bar\nn'$ 
by their unbarred values (these will differ
only by terms of order $v/c$ and the difference when multiplied by 
$i_{2m}$ will not contribute to the first order).

Finally, we also need to evaluate $d\tau/d\eta$. 
Since $A_1$ and $A_2$ are already of first order, we 
need to evaluate this term only to zero'th order,
to write down the equation for the perturbed brightness. We have 
$d\tau/d\eta = \bar u^0/u^0 = a$ to the leading order.
The perturbed brightness equation \eq{final}, given in the main text,
is got from \Eq{btli}, \Eq{ifrt} and \Eq{atwo}.

\section{Silk damping: details}
\label{silk}

We have given in the main text the iterative solution to
\Eq{finit} to the first order in $L_\gamma$. To derive Silk
damping one needs to go to the second order iteration,
\EQA
i^{(2)} &=& i^{(0)} - L_\gamma 
\left[\frac{\partial i^{(1)}}{\partial \eta} 
+\nn\cdot\nab i^{(1)} + 8\nn\cdot\nab\phi \right] \nonumber \\
&=& i^{(1)} + L_\gamma^2 \left[\frac{\partial}{ \partial \eta} 
+\nn\cdot\nab \right]\left[\frac{\partial i^{(0)}}{\partial \eta}     
+\nn\cdot\nab i^{(0)} + 8\nn\cdot\nab\phi \right]
\label{ftwo}
\ENA
As mentioned in the text, we neglect the anisotropy of
the Thomson scattering, and also the effects of
the gravitational potential $\phi$. 
Taking the zeroth moment of \Eq{ftwo} , we get
\EQ
i_0 = i_0 - L_\gamma \left[\frac{\partial i_0}{\partial \eta} + 
\frac{4}{3} \nab\cdot\vv \right ] 
+  L_\gamma^2  \left [
\frac{\partial^2 i_0}{\partial \eta^2} + 
\frac{8}{3} \frac{\partial(\nab\cdot\vv)}{\partial \eta}
+ \frac{1}{3} \nab^2 i_0 \right ]
\label{iotwo}
\EN
So to the next order in $L_\gamma$, \Eq{coni} is modified to 
\EQ
\frac{\partial i_0}{\partial \eta} + \frac{4}{3} 
\nab\cdot\vv = L_\gamma \left [
\frac{4}{3} \frac{\partial(\nab\cdot\vv)}{\partial \eta}
+ \frac{1}{3} \nab^2 i_0 \right ]  
\label{conin}
\EN
Similarly, taking the first moment of \Eq{ftwo} the Euler equation 
\Eq{EULR} gets modified to 
\EQ
\left[\rho_B + \frac{4}{3} \rho_R \right]
\frac{\partial \vv}{ \partial \eta}
= -\nab \left(\frac{\rho_R i_0}{3}\right) 
+\rho_R L_\gamma \left [ 
\frac{4}{3} \frac{\partial^2 \vv }{\partial \eta^2}  +
\frac{2}{3} \nab \left(\frac{\partial i_0}{\partial \eta}\right) 
+ \frac{8}{15} \nab(\nab\cdot\vv)
+ \frac{4}{15} \nab^2\vv \right ]
\label{eulra}
\EN
Here we have used the relation
$ \int (d\Omega/4\pi) n_in_jn_kn_l =
[ \delta_{ij}\delta_{kl} +
 \delta_{ik}\delta_{jl} +\delta_{il}\delta_{kj} ]/15 $.
(This can be written down from symmetry and the coefficients
and its amplitude fixed by contracting 
over any two indices).
We have also neglected the baryonic pressure 
compared to the radiation pressure.
Equations \eq{conin} and \eq{eulra} form a  pair of 
linear coupled equations for the 
the perturbations in radiation density $i_0$ and matter velocity 
$\vv $. 
Assuming that the rate of variation of
of the co-efficients of various terms, due to Hubble expansion is small
(compared to $kc_s$), one can use the WKBJ approximation,
to derive the dispersion relation for the baryon-radiation 
acoustic oscillations.  

Consider therefore a plane wave solution of the form 
\EQ
\vv  = \VV \exp(\ii\kk\cdot\xx + \int \Gamma d\eta) ; \quad
i= I \exp(\ii\kk\cdot\xx + \int \Gamma d\eta)
\label{forierb}
\EN
Let us also look at longitudinal waves with $\kk$ parallel to 
$\VV $. Infact, taking the divergence of 
\Eq{eulra} one can see that these modes are completely 
decoupled from the rotational modes.
To leading order one gets from
\Eq{conin} and \Eq{eulra}, a dispersion relation which is
a cubic equation for $\Gamma$, 
\EQ
- L_\gamma \Gamma^3 + b\Gamma^2 + b\Gamma \frac{k^2 L_\gamma}{3} 
\left[1 - \frac{6}{5b}\right] 
+ \frac{k^2}{3} = 0
\label{cub}
\EN
which can be solved iteratively. Here we have defined 
$b = (1 + 3\rho_B/4\rho_R) = 1+R$.
To the lowest order we get $\Gamma = \pm i (k/\sqrt{3b})$. So to the 
zeroth order the dispersion relation is that of a sound (pressure) 
wave in the baryon-photon fluid, with an effective sound speed 
$c_s =(1/\sqrt{3b})$. 
Consider the effects of terms proportional to $L_\gamma$. 
Since the $\Gamma^3$
term is already multiplied by $L_\gamma$ 
we can use the lowest order solution
to write $-L_\gamma\Gamma^3 =  -L_\gamma (-k^2/3b)\Gamma$. This reduces 
the cubic equation to the quadratic equation 
\EQ
\Gamma^2 + \Gamma \frac{k^2 L_\gamma}{3} 
\left[1 - \frac{6}{5b} + \frac{1}{b^2}\right] 
+ \frac{k^2}{3b} = 0
\label{dispb}
\EN
whose solution to first order in $L_\gamma$ 
is \Eq{disp} given in the main text.

\begin{figure}[t!]\begin{center}
\includegraphics[width=.9\textwidth]{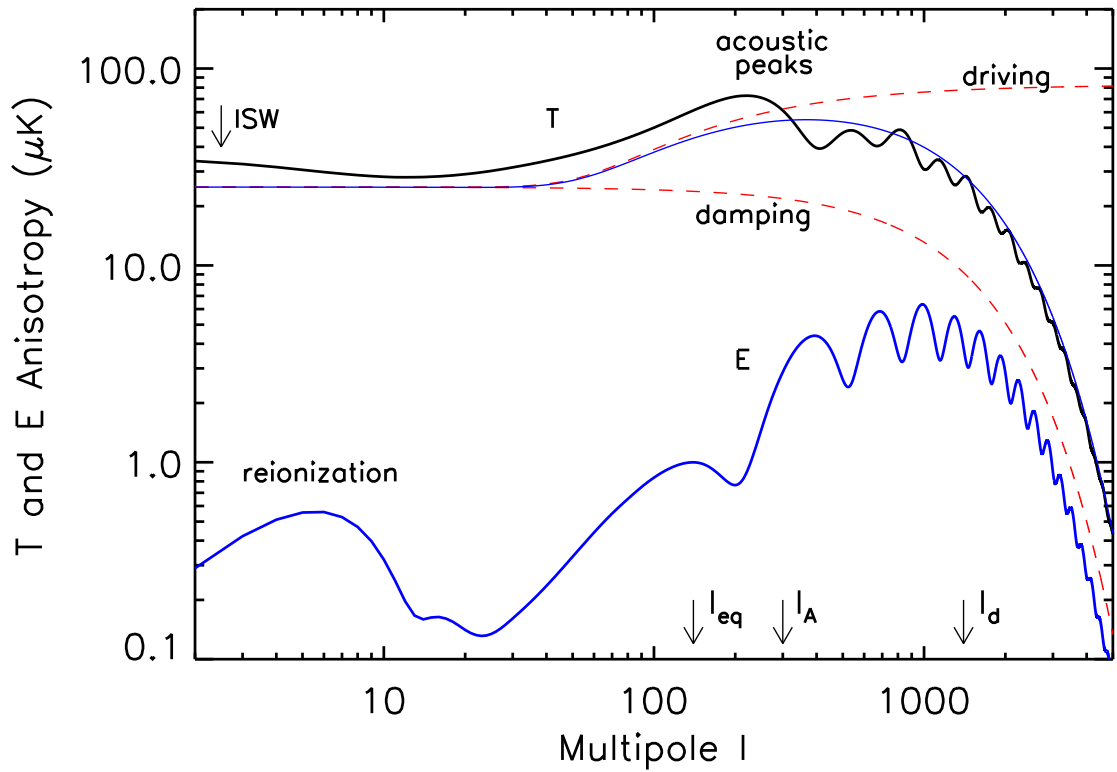}
\end{center}\caption[]{Temperture and E-type 
polarization anisotropies versus the multipole number,
for a flat $\Lambda$CDM model, consistent with WMAP and
computed using CMBFAST \cite{cmbfast}.
The parameters are: $\Omega_b =0.046, \Omega_m = 0.27,
\Omega_\Lambda = 0.73, h=0.72, n_s=0.99$. Some of the effects
discussed in the text are marked in the figure, as well as the
location of different characteristic scales $l_{eq}$, $l_A$ and
$l_D$. Radiation driving leads to the rise of the temperature
anisotropy above $l_{eq}$, while Silk damping and the damping
due the finite thickness of the LSS causes the amplitude 
to fall for $l > l_D$. The forms for these envelopes
are taken from \cite{hu_white97}. The ISW effect is important
at small $l$. The early ISW effect is important
around $l_{eq}$ and is one reason for the first peak's shift to
$l_1 < l_A$. The polarization rises as $(l/l_D)$ in the tight coupling
limit, due to the small quadrupole source, and is also
damped for $l > l_D$. The effect of reionization is to
cause the rise of the polarization signals at low $l$. 
The figure is inspired by a similar figure in Hu \cite{hurev}.}
\label{ptriple}\end{figure}

\begin{figure}[t!]\begin{center}
\includegraphics[width=0.9\textwidth]{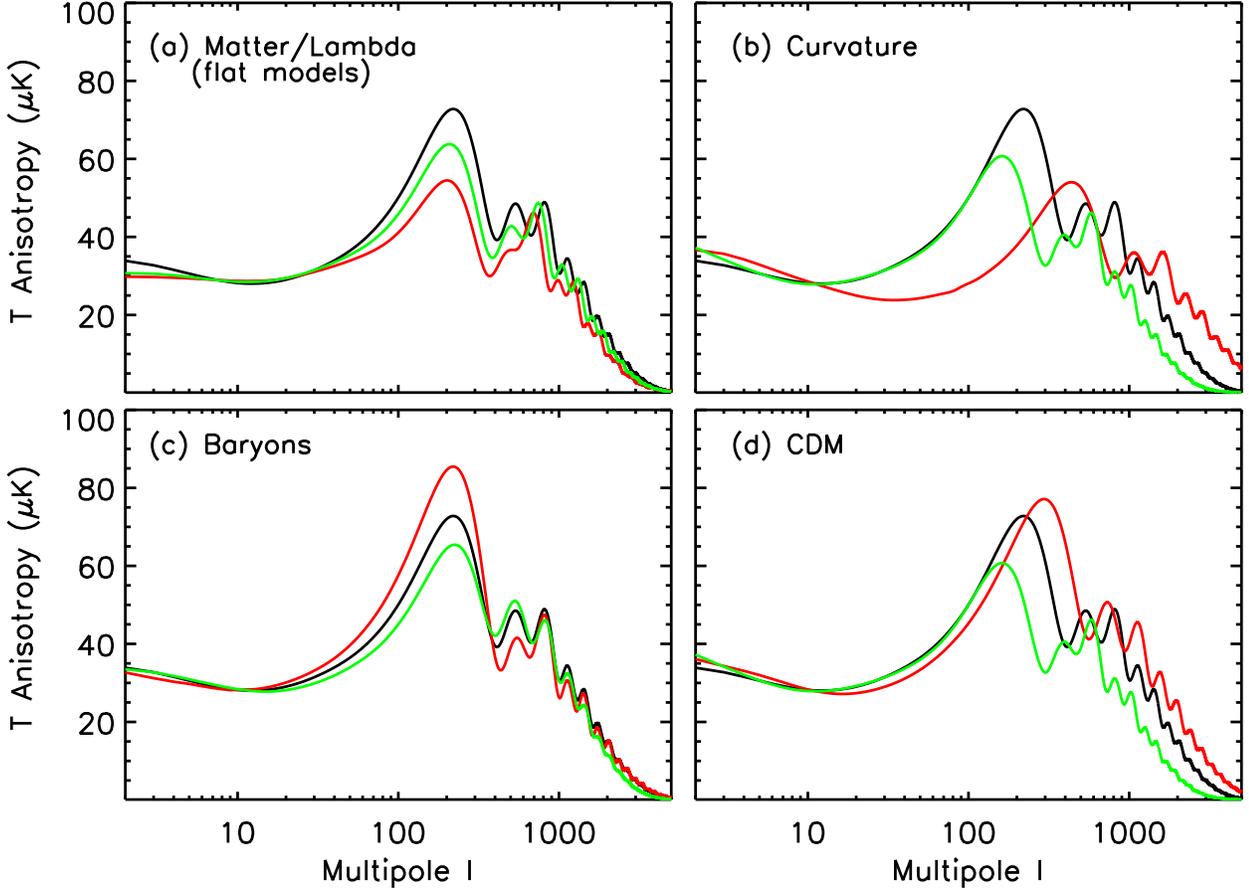}
\end{center}\caption[]{The figure illustrates the sensitivity 
of the temperature anisotropy
to changes in various parameters, about the standard $\Lambda$CDM
model of Figure 1. The black (dark) line is always the standard
$\Lambda$CDM model. Figure (a): Shows sensitivity to matter/lambda
densities for flat models. The three plots
with decreasing peak amplitudes result from 
increasing $\Omega_m$ to $0.5$ and $1$, keeping total $\Omega_T=1$.
Figure (b): Here the effect of total curvature on peak location
is illustrated. The plot with the smallest $l$ value for the first peak
has $\Omega_m=0.448$, $\Omega_\Lambda=0.73$, while that
with the largest is for an open universe, with
$\Omega_m =0.27$ and $\Omega_\Lambda=0$. Figure (c): Shows the
sensitivity to the baryon density. For the three plots
with decreasing first peak amplitudes, one has
$\Omega_b = 0.092$, $0.046$ and $0.023$ respectively. 
Figure (d): The sensitivity to the CDM density parameter.
This is $0.112$, $0.224$ and $0.448$
for the three plots, where the first peak progressively 
shifts to lower $l$ values.
}\label{pt}\end{figure}

\end{document}